\documentclass[a4paper,12pt]{iopart}

\usepackage{graphicx}

\usepackage{amssymb}

\usepackage[english]{babel}
\usepackage{color}

\newcommand{\ket}[1]{\bigl|#1\bigr\rangle}
\newcommand{\bra}[1]{\bigl\langle#1\bigr|}

\newcommand{\proj}[2]{\ket{#1}\bra{#2}}

\newcommand{\text}[1]{\mathrm{#1}}
\newcommand{\eqref}[1]{\eref{#1}}

\begin{document}

\title{Motional effects on the efficiency of excitation transfer}

\author{Ali Asadian$^{1,2}$, Markus Tiersch$^{1,2}$, Gian Giacomo Guerreschi$^{1,2}$, Jianming Cai$^{1,2}$, Sandu Popescu$^{3,4}$, and Hans J. Briegel$^{1,2}$}

\address{$^1$
Institut f{\"u}r Theoretische Physik, Universit{\"a}t Innsbruck,
Technikerstra\ss{}e 25,
A-6020 Innsbruck, Austria}
\address{$^2$
Institut f\"ur Quantenoptik und Quanteninformation
der \"Osterreichischen Akademie der Wissenschaften,
Innsbruck, Austria}
\address{$^3$
H.H. Wills Physics Laboratory, University of Bristol,
Tyndall Avenue, Bristol BS8 1TL, U.K.
}
\address{$^4$
Hewlett-Packard Laboratories, Stoke Gifford,
Bristol BS12 6QZ, U.K.
}
\ead{hans.briegel@uibk.ac.at}


\begin{abstract}
Energy transfer plays a vital role in many natural and technological processes.
In this work, we study the effects of mechanical motion on the excitation transfer through a chain of interacting molecules with application to biological scenarios of transfer processes.
Our investigation demonstrates that, for various types of mechanical oscillations, the transfer efficiency is significantly enhanced over that of comparable static configurations.
This enhancement is a genuine quantum signature, and requires the collaborative interplay between the quantum-coherent evolution of the excitation and the mechanical motion of the molecules;
it has no analogue in the classical incoherent energy transfer.
This effect may not only occur naturally, but it could be exploited in artificially designed systems to optimize transport processes.
As an application, we discuss a simple and hence robust control technique.
\end{abstract}


\maketitle

\section{Introduction}

Transport phenomena can be safely enumerated amongst the most important physical processes occurring in both natural and artificial systems, and therefore the transport of charges or energy plays a central role in many scientific disciplines.
Particularly important examples are the life-enabling transport processes in the molecular machinery of biological systems, which take place at scales ranging from a few atoms to large macro-molecular structures such as photosynthetic complexes.
Pioneering studies of energy transfer, both theoretical and experimental, have already appeared in the early stages of the development of quantum mechanics~\cite{perrin32}, and this subject has been of interest ever since.
Recent experimental investigations reveal the presence of long-lived quantum coherence during charge or energy transport processes in photosynthetic complexes~\cite{Brixner05,Engel07,Lee07,Ishizaki09}, charge transport through DNA~\cite{Giese01}, and in polymers~\cite{Collini09}, even at physiological temperatures~\cite{Engel10,Collini10}.
These findings raise the following question: To what extent may a fully quantum-coherent nature of the transport processes (rather than a classically incoherent, diffuse hopping), and perhaps even the presence of stronger quantum signatures such as quantum entanglement~\cite{Horodecki09} be responsible for the remarkable efficiencies that we witness in these systems~\cite{Cai08,Sarovar09,Caruso09a,Scholak09}.

At room temperature, motion of the underlying atomic and molecular structures, along which transport happens, is a ubiquitous phenomenon.
Perhaps more important for living systems is the origin of molecular motion due to driving.
The relevant time-scales range from a few femtoseconds for a stretching mode oscillation of two covalently bound atoms to several picoseconds for the collective modes of larger molecules, and beyond for the conformational changes of proteins~\cite{Schotte03,Frauenfelder09}.
If the transport process and the underlying structure evolve on comparable time-scales, their dynamics do not separate, and we must expect an interplay between both that may affect the transport efficiency.

In the present work, we treat an abstract transport model in which the motion of  the molecule backbone structure is classical, and focus on the quantum-coherent transport that takes place on this moving structure.
The quantum degrees of freedom that are carried by the molecular backbone will thus inherit time-dependent features from the classical motion.
We show how such a time dependence can actively drive the coherent excitation transfer and thereby increase its efficiency beyond any comparable static scenario.
The presence of this enhancement is a genuine quantum effect as it does not occur in a classical diffusive hopping transfer.
With this we identify and quantify the motion-induced quantum enhancement of the quantum-coherent transport.

It is known that dynamical structural changes in proteins affect transport processes in various ways.
Environmental noise induced by thermal motion, and specifically dephasing, may lead to an enhancement of the efficiency of transport through a network of sites~\cite{Balabin00,Plenio08,Mohensi08,Caruso09,Semiao09,Chin09}.
There, the generic idea is that dephasing can overcome the trapping of energy due to a disorder of the local energies, or due to destructive interference along different paths through the network (similar as destructive interference in a precisely aligned interferometer).
As an example, such a dephasing process can result from coupling the local energies to the collective vibrations of the network.
The transport process itself may, however, induce a rearrangement of the surrounding molecular structure that acts back on the transport process~\cite{Schulten91,Reddy93,Vos94,Goushcha99,Kriegl04,Wang06}, and may induce a directionality, for example.
In this case, protein motion and transport processes are coordinated as opposed to random changes due to thermal motion.
Indeed, such triggered, long-lived coherent motion of protein nuclei has been found to exist~\cite{Vos93,Vos00}, even at room temperature.

Here, we choose an approach consistent with the latter observation and treat the molecule motion as coordinated with the transfer process rather than as a noise source.
Therefore, in our case the enhancement of transport efficiency has nothing to do with dephasing, but arises from driving, i.e.\ a {\em constructive} interplay between the structural changes of the transport network and the quantum-coherent hopping between the sites.
Since the underling physical mechanisms of single-electron, hole, proton, or spin transport have an identical formal modeling, the results of this study can be straightforwardly applied to the corresponding parameter regimes of these systems as well.
Furthermore, we demonstrate simple and therefore robust control techniques to induce or enhance these effects in linear systems.

\paragraph{}%
In the present work, we shall deal with a rather abstract model to investigate the effects of motion on transport efficiencies.
Nevertheless, several biological applications are conceivable that largely resemble our chosen parameter regimes, and show features that agree with the assumptions underlying our model.

In proteins, it is suggested that energy transport plays a crucial role for conformational changes~\cite{CruzeiroReview,Xie02}.
For the protein to change its shape, and perform its function, it is often necessary to overcome an energetic barrier.
It is hypothesized that the necessary energy needs first to be available as vibrational energy in the protein, and that it is transported from a site where it is provided by ATP to the region within the protein where it is consumed by a conformational change.
A type of secondary structure in proteins, in which the chain of amino-acids forms an $\alpha$-helix, is suggested to provide such a transport pathway for vibrational energy~\cite{CruzeiroReview,Xie02}.
Although the sites that carry the excitation in form of a localized molecular vibration (the amide-I oscillator is predominantly a \mbox{$\mathsf{C=O}$} stretching mode oscillation) are arranged in a helix, the structure is essentially linear.
In this scenario, energy transport has been investigated under the name of the Davydov-Scott-model~\cite{CruzeiroReview,Scott92,Davydov}, to a greater degree of realism than we aim to achieve here, and with an emphasis on localization phenomena of excitations on the $\alpha$-helix due to motion~\cite{Cruz05,Tsivlin05,Tsivlin07}.

For photosynthesis in green sulfur bacteria, light is harvested by a large antenna complex, and the excitation is subsequently transferred via the well-studied Fenna-Matthews-Olson (FMO) protein, that thus functions as a wire, to the reaction center where the excitation energy is converted to chemical energy, for a review see~\cite{Cheng09} and references therein.
Within a single unit of the FMO trimer, the excitation has been shown to propagate quantum-coherently via seven bacteriochlorophyll pigments, entering and exiting at specific pigments~\cite{Brixner05,Engel07,Ishizaki09,Engel10}.
In the reaction center, an electron is then quantum-coherently transferred via several sites.
This electron transfer is accompanied by a structural changes in form of low frequency vibrations of the surrounding protein environment on the same time scale~\cite{Vos93,Vos00,Lee07}.

\paragraph*{}
The paper is organized as follows.
In section~\ref{basic setting}, we introduce the formal model, and start our investigation with the simplest possible case of energy transfer, namely the dimer model that consists of only two molecules.
This serves as a reference system to build intuition of how mechanical oscillations can enhance the energy transfer.
By comparing quantum excitation transfer with the classical F\"{o}rster theory of energy transfer, we explain how mechanical motion can constructively cooperate with the quantum dynamics of the excitation.
In section~\ref{standing oscillations}, we generalize this model to moving systems of many molecules, where we investigate motion modeled by their normal modes and the resulting effects on excitation transfer.
Section~\ref{guided transfer} extends the study of motion of a chain towards a scheme in which the interaction strength between sites is modulated by a pulse.

\section{Transport Model}
\label{basic setting}

\begin{figure}
\begin{minipage}[c]{0.7\textwidth}
\centering
\includegraphics{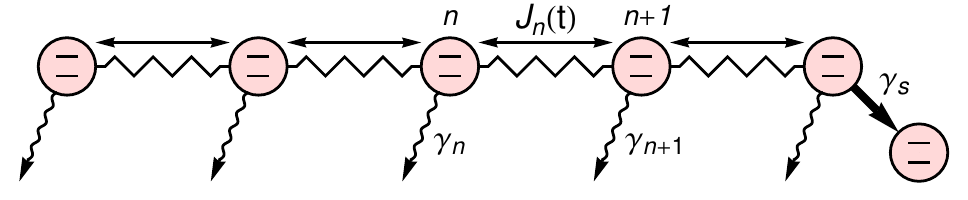}%
\end{minipage}%
\begin{minipage}[c]{0.3\textwidth}
\centering
\includegraphics[height=4.5cm]{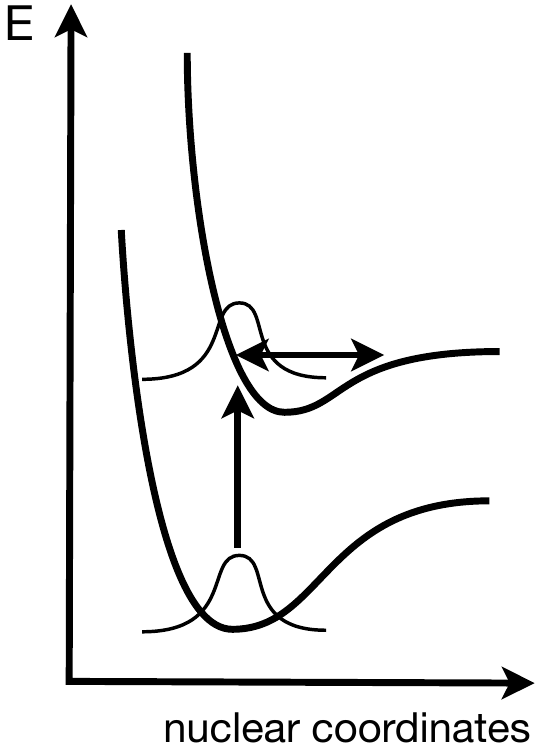}%
\end{minipage}
\caption{Left: Schematic transport model of next-neighbor coupled chain with local dissipation and the sink attached to the last site. Right:~Schematic energy diagram of two electronic states before and after excitation showing how excitation can induce motion of the nuclei that is coordinated with the launch of the excitation propagation in the complex~\cite{Vos94}.}
\label{chainmodel}
\end{figure}

Let us first set the formalism to describe excitation transport.
Here, we limit ourselves to a linear chain of $N$ sites, which represents an array of molecules, e.g.~an $\alpha$-helix, among which an excitation can be exchanged due to dipole-dipole couplings between the molecules.
When the probability of an excitation being present in the system is low, we can restrict our analysis to the single-excitation subspace.
Furthermore, we assume for simplicity that the molecules are sufficiently distant from each other such that their interaction can be reduced to the dominant interaction between nearest neighbors.
Generally, however, this is not the necessarily the case, as in the FMO complex, for example.
With these assumptions, the Hamiltonian of the system can be written as the tight-binding Hamiltonian of an interacting $N$-body system in the single-excitation manifold ($\hbar=1$),
\begin{equation}
\label{Hamiltonian}
 H=\sum_{n=1}^{N}\varepsilon_{n}\proj{n}{n} + \sum_{n=1}^{N-1}J_n \Big( \proj{n}{n+1}+\proj{n+1}{n} \Big) ,
\end{equation}
where $\ket{n}$ denotes the state with the excitation at the $n$-th site having energy $\varepsilon_{n}$, and $J_n$ is the coupling strength between the $n$-th and $(n+1)$-th molecule, as depicted in Figure~\ref{chainmodel}~(left).

We assume the initial state of the system at time $t=0$ to be a single excitation localized at the first site (left end of the chain in figure~\ref{chainmodel}), i.e.~$\rho(0)=\proj{1}{1}$.
This is in accordance with the idea that the excitation enters the transport network at specific sites as verified in the FMO complex~\cite{Wen09}, or provided locally by ATP in the $\alpha$-helix scenario.
For spectroscopic investigations, fs-pulses of the right wave length are able to excite specific sites with in a complex of chromophores as done in the FMO kind~\cite{Brixner05}.

The motion of molecules will impose a time dependence onto the Hamiltonian~\eqref{Hamiltonian}.
We treat the spatial positions of the chain constituents as classical quantities that follow well-defined trajectories~\cite{Cai08}.
This semi-quantal approximation holds if the involved molecules are too large to show their quantum behavior, i.e.\ if the uncertainty with which the position of a molecule can be determined due to the classical thermal fluctuations is larger than the quantum width of the associated wave function in the position representation (see also~\cite{Cruz97}).
For a molecule of mass $M$ that is attached to a spring with spring constant $K$, and that is in contact with an environment at temperature $T$, the uncertainty of its position due to thermal fluctuations is $\Delta x_{th}^2 = 2 k_{B} T/K$, while the width of the (ground state) wave function is $\Delta x_{qu}^2 = 2 \hbar/\sqrt{M K}$, with $k_{B}$ being the Boltzmann constant, and $\hbar$ the Planck constant.
Therefore, $\Delta x_{th}>\Delta x_{qu}$ if the temperature is high enough, i.e.\ if $k_B T>\hbar\sqrt{K/M}$, and we can focus on classical motion of the excitation carriers.
For example, the typical critical temperature for an $\alpha$-helix structure is 60\,K (as obtained for the mass and spring constant in the effective one-dimensional model: $K\simeq39-58.5$\,N/m, $M\simeq5.7 \times 10^{-25}$\,kg~\cite{Scott92}) which is certainly much below room temperature.

Motion of the individual molecules changes their relative distance in time and thereby modulates the distance-dependent dipolar coupling.
At the same time, deformation of the molecules leads to a change in dipole moments and thus also induces a time-dependence of the coupling.
The latter will be dominant whenever the molecules are tightly embedded in a protein scaffold such as in the FMO complex.
For our purposes, and in order to demonstrate the effect, we only focus on a modulation of $J_n$ due to the change in distance, since we expect a change of dipoles in strength or direction to yield similar results.
For concreteness, we choose a simple form of motion where neighboring molecules change their distance periodically, which holds for small amplitudes in the harmonic regime.
We will later see that for typical times that the excitation spends in the complex only few molecule oscillation periods at the relevant frequencies suffice such that we essentially disregard damping and dephasing of the molecule motion.
For times short compared to the damping of the motion, the distance between site $n$ and $n+1$ is then given by
\begin{equation}
\label{distance}
d_n(t)=d_0 - \big[u_n(t)-u_{n+1}(t)\big]=d_{0} \left[ 1-2a_n\sin(\omega t+\phi_n) \right],
\end{equation}
where $u_n(t)$ is the displacement of the $n$-th molecule, $d_0$ is the equilibrium distance between two neighboring sites, and $a_n$ is the individual sites' relative amplitude of oscillation when they move with opposite phase around their equilibrium position.
This affects the dipolar coupling strengths $J_n$ according to
\begin{equation}
\label{eq:coupling}
J_n(t)=\frac{\tilde{J}_{0}}{[d_n(t)]^3}
=\frac{J_{0}}{[1-2a_n\sin(\omega t+\phi_n)]^3},
\end{equation}
where $\tilde{J}_0$ contains the dipole moments (here assumed to be constant) and physical constants, and we define $J_0=\tilde{J}_0/d_0^3$, which has the unit of an energy.
The extrema of the coupling are $J_{n,\text{min}}$ when the molecules are farthest apart from each other, that is, when $d_n=d_{0}(1+2a_n)$, and $J_{n,\text{max}}$ when their distance reaches the minimum value of $d_n=d_{0}(1-2a_n)$.
The deterministic motion of the molecules imposed by~\eqref{distance} is for a given amplitude, frequency, and phase synchronized with the propagation of the excitation.
This is in contrast to a stochastic motion of the molecules as caused by thermal fluctuations, for example.
It is, however, conceivable that upon the arrival on the excitation in the complex the electron configuration changes and nuclei begin to move in this new potential due to the Franck-Condon principle as shown in figure~\ref{chainmodel}~(right)~\cite{Vos94}.
The wave packet that describes the nuclear motion has been observed to exhibit surprisingly long coherence times in reaction center proteins~\cite{Vos93}, which is consistent with our ansatz to treat positions of the sites classically, and in concert with the initial excitation.

Other influences that motion might have on the dynamics of the transport include the detuning of the excitation energies of the individual sites.
Here, we will mostly ignore the such arising disorder and localization effects.
In a realistic biological context such as excitation transport in light harvesting complexes, these effect are usually not negligible~\cite{Damjanovic02}.

\paragraph{}
In addition to the Hamiltonian evolution, the excitation dynamics takes place in an open quantum system and therefore also suffers from dissipation.
We describe the loss of the energy excitation due to dissipation in Markov approximation by the following Lindblad super-operator:
\begin{equation}
L_\text{diss}\rho=\sum_{n=1}^{N}\gamma_{n}\big[2\sigma_{n}^{-}\rho\sigma_{n}^{+}-\{\sigma_{n}^{+}\sigma_{n}^{-},\rho\}\big].
\end{equation}
where $\sigma_{n}^{+}$ ($\sigma_{n}^{-}$) is the creation (annihilation) operator of the excitation at site $n$, and $\{A,B\}\equiv AB+BA$.
Since realistically each molecule feels a different environment, the local dissipation rates $\gamma_n$ are different in general.
For simplicity and due to so far unavailable experimental data, however, we will later assume them to be equal.

In order to measure how much of the excitation energy is transferred along the chain (and not lost due to dissipation), we introduce an additional site, the sink, representing the final ($N+1$)-th trapping site that resembles a reaction center, for example.
The sink is populated via irreversible decay of excitation from the last site, ~$N$.
This approach for quantifying the energy transfer efficiency follows Ref.~\cite{Plenio08}, and is formally implemented by adding the Lindblad operator
\begin{equation}
L_\text{sink}\rho=\gamma_{S}[2\sigma_{N+1}^{+}\sigma_{N}^{-}\rho\sigma_{N}^{+}\sigma_{N+1}^{-}-\{\sigma_{N}^{+}\sigma_{N+1}^{-}\sigma_{N+1}^{+}\sigma_{N}^{-},\rho\}]
\end{equation}
to the master equation, where $\gamma_{S}$ denotes absorption rate of the sink.

In order to calculate the efficiency of the energy transfer given by the asymptotic population of the sink, it is necessary to integrate the following master equation $(\hbar=1)$:
\begin{equation}
\label{mastereq}
\frac{\partial\rho}{\partial t}=L\rho=i[\rho,H(t)]+L_\text{diss}\rho+L_\text{sink}\rho.
\end{equation}
Since we are interested in the asymptotic population of the sink, which is our figure of merit for the transport efficiency, we search for long-time solutions of the above equation.
Asymptotically in time, the excitation will be either lost to the environment due to $L_\text{diss}$ or, ideally, trapped in the sink due to $L_\text{sink}$.
Note that it therefore does not suffice to merely search for the stationary state, which is characterized by the ground state of the $N$ sites: it thus contains no information about how likely the excitation is trapped in the sink instead of being lost to the environment.

In numerical simulations, we will usually consider uniform local energies $\varepsilon_{n}=\varepsilon$ unless otherwise noted.
Additional effects that arise from static disorder or detuning of the local energies due to the coupling to the nuclear motion are outside the focus of the present work, and are thereby mostly neglected.
Henceforth, all energies, time-scales, and rates will be expressed in units of $J_0$, and thereby we effectively set $J_0=1$.

\subsection*{Dimer Model}

In the remainder of this section, we study the case of a short chain composed of only $N=2$ interacting sites, the dimer, plus the sink in a local dissipative environment for $\varepsilon_1=\varepsilon_2\equiv\varepsilon$.
Although simple, this toy model exhibits all the essential features of larger systems, and gives us the possibility of guiding our intuition towards the mechanisms that underlie the transport processes.

For a time-\emph{independent} coupling $J$, the excitation coherently oscillates between both sites at a constant Rabi frequency until it is either lost into the environment or absorbed into the sink, when it is at site~2.
The asymptotic sink population is given, in this static situation, by~\cite{Plenio08}
\begin{equation}
P_\text{sink}^\text{st}(J)=\frac{\gamma_{S}J^{2}}{(2\gamma+\gamma_S)[\gamma(\gamma+\gamma_S)+J^2]}
\end{equation}
where $\gamma_1=\gamma_2\equiv\gamma$ is the rate of dissipation into the environment at each site, and the superscript ``$\text{st}$'' indicates the static, i.e.\ non-oscillatory, case with constant $J$.

\begin{figure}
\centering
\includegraphics[scale=0.8]{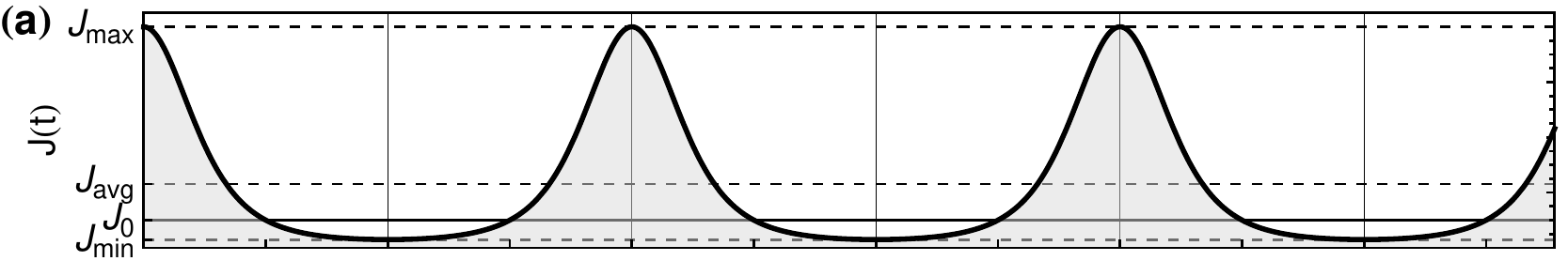}
\includegraphics[scale=0.8]{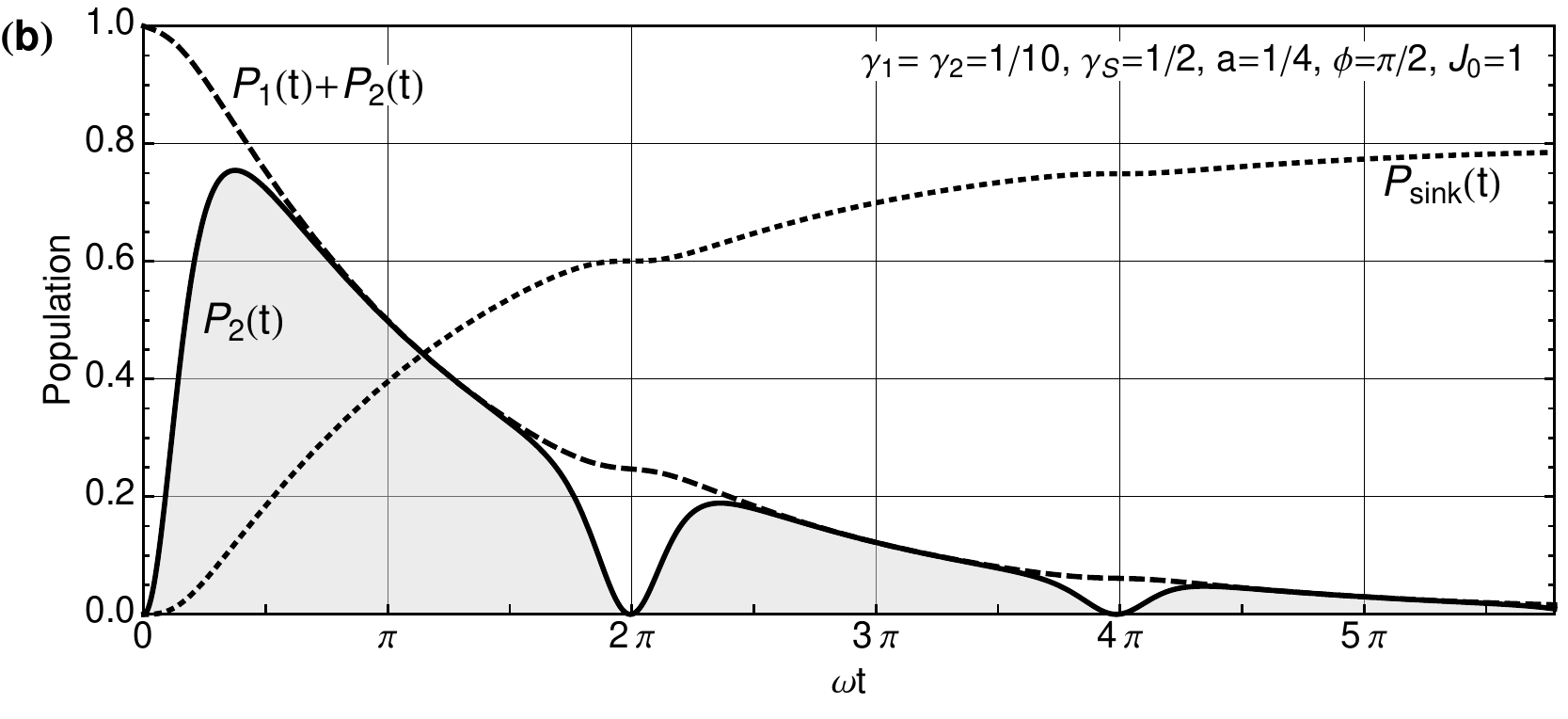}
\caption{(a) Coupling strength. (b) Population dynamics for coherent transfer in a dimer molecule oscillating at the optimal frequency.
Curves represent the population of site~$2$ (solid), the sum of the populations in site $1$ and $2$ (dashed), and the population of the sink (dotted).
The time axis is rescaled using the oscillation frequency $\omega=4.54$, which corresponds to the maximum in figure~\ref{freqEstimate}(a).}
\label{2sites_optimalFreq}
\end{figure}

In the case of a time-\emph{dependent} coupling $J(t)$, the master equation can be straightforwardly integrated if the combined dissipation rates at both sites are equal, i.e.\ for $\gamma_1=\gamma_2+\gamma_S\equiv\Gamma$, and $\varepsilon_1=\varepsilon_2$ as above. With this choice, the populations of the respective sites read:
\begin{eqnarray}
\label{sitePop1}
P_1(t) =\cos^2\left(\int_0^t J(t^\prime) dt^\prime\right) e^{-2\Gamma t},
\qquad \text{and} \\
\label{sitePop2}
P_2(t) =\sin^2\left(\int_0^t J(t^\prime) dt^\prime\right) e^{-2\Gamma t}.
\end{eqnarray}
The sink population is finally obtained by integration: $P_\text{sink}(t)=2\gamma_S \int_0^t P_2(t) dt$.

Since the phase of the Rabi oscillation is obtained by integration over the coupling strength in \eqref{sitePop1} and \eqref{sitePop2}, we expect a faster propagation of the excitation between sites whenever the coupling strength is stronger, i.e.\ when the two sites are closer to one another.
Similarly, we witness a slower excitation propagation (or even an almost static situation) during times of weak coupling, when the sites are distant.
For a relative amplitude of $a\equiv a_1=1/4$, figure~\ref{2sites_optimalFreq}(a) shows the time-dependence of the interaction strength, which exhibits (for the chosen $a$) short times of strong interaction, separated by longer durations of weak coupling.
This asymmetry in the coupling strength is due to the disproportional dependence of $J(t)$ on the relative distance of the sites, cf.~\eqref{eq:coupling}.

For the right ratio of Rabi frequency and mechanical oscillation frequency $\omega$, the narrow peaks of strong interaction, when the sites are close, can be used to quickly transfer the excitation between the sites, whereas the longer ``valleys'' of low interaction strengths and slow Rabi oscillation are used to effectively lock the excitation on one site, cf.~figure~\ref{2sites_optimalFreq}(b).
Using this quick-transfer-and-locking strategy, we can estimate parameters that allow us to maximize the duration during which the excitation is located at the second site, and thereby most efficiently exposed to the decay into the sink.
Starting with the initial spatial position given by the phase $\phi\equiv\phi_1=\pi/2$ such that the sites are closest, and the interaction is strongest, we need about the first 1/4th of the mechanical oscillation cycle, when the interaction is still sufficiently strong, to transfer the excitation from site 1 to site 2 (in figure~\ref{2sites_optimalFreq}(b) $P_2$ approaches the total population of both sides).
During the next half cycle of the mechanical oscillation, when the interaction strength is low, the site population does not change significantly.
This effectively amounts to a locking of the excitation at site 2.
The following half-cycle, during which the interaction is strongest again, then needs to last long enough for the excitation to move again to site 1 and back.

\begin{figure}
\centering
\includegraphics[width=0.6786\textwidth]{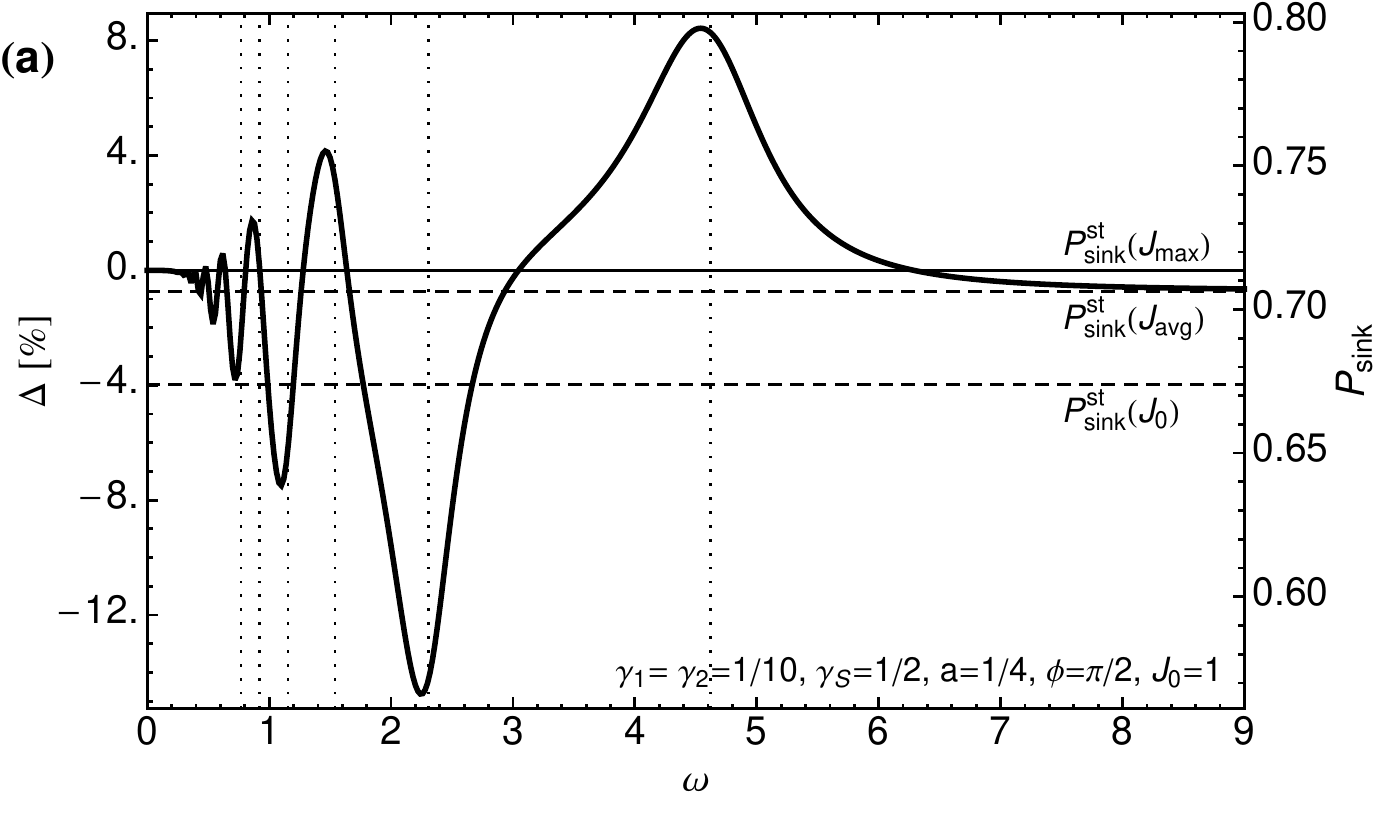}
\hfill
\includegraphics[width=0.2714\textwidth]{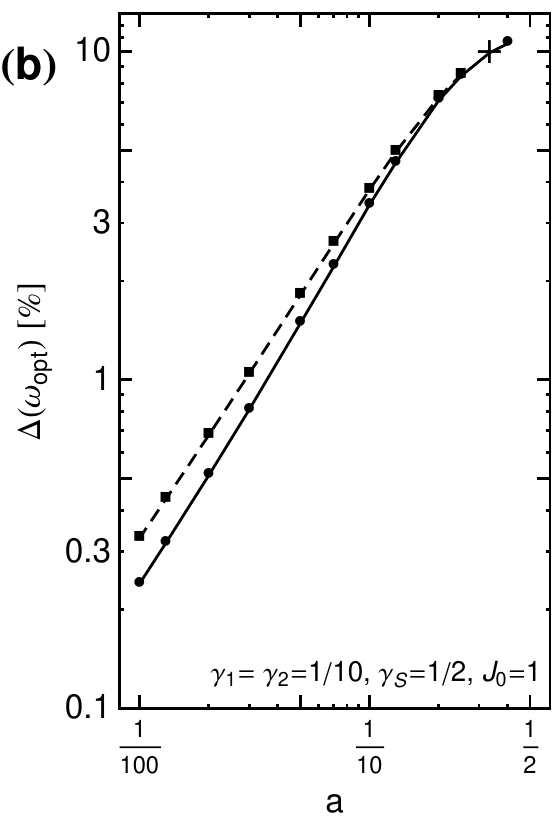}
\caption{Asymptotic sink population dependent on the mechanical oscillation frequency $\omega$. (a) Initial phase of the mechanical oscillation is $\phi=\pi/2$, i.e.\ the first site of the chain is excited when the sites are closest. Estimated frequencies according to \eqref{freqEstimateMax} and \eqref{freqEstimateMin} for extremal sink population are indicated by vertical lines.
Horizontal lines indicate the asymptotic sink population for static couplings.
For asymptotically large $\omega$, the curve approaches the sink population $P_\text{sink}^\text{st}(J_\text{avg})$.
(b)~Amplitude dependence of the maximal enhancement with an initial phase $\phi=\pi/2$ (solid) and the optimal initial phase (dashed) shows an algebraic decrease for small amplitudes in a double logarithmic plot. The cross marks the amplitude used for the other plots.
}
\label{freqEstimate}
\end{figure}

Comparing with \eqref{sitePop1} and \eqref{sitePop2}, the above translates into the requirement that \mbox{$\int_0^{T/4}J(t^\prime) \rmd t^\prime\simeq\pi/2$} for the first transfer to site~2, where $T=2\pi/\omega$ is the period of the mechanical oscillation.
For a strong interaction modulation as in figure~\ref{2sites_optimalFreq}(a), we neglect the coupling strength when the two sites are distant, and the integral can be approximated by $J_\text{avg}\pi/\omega$, where \mbox{$J_\text{avg}=J_0(1+2a^2)/(1-4a^2)^{5/2}$} denotes the time-averaged coupling strength.
For the above situation, we thereby obtain an optimal frequency of $\omega \simeq 2 J_\text{avg}$.
Higher-order maxima are obtained by adding full Rabi cycles to a transfer step, i.e.\
\begin{equation}
\label{freqEstimateMax}
\int_0^{T/4} J(t^\prime) \rmd t^\prime \simeq \frac{\pi}{2}+m\pi
\quad \textrm{and} \quad
\omega \simeq \frac{2J_\text{avg}}{2m+1}, \quad m=0,1,2,\ldots.
\end{equation}
In contrast, for a locking of the excitation at site 1, i.e.\ for a minimum exposure to the sink decay, an analogous argument yields the minima of the transfer efficiency:
\begin{equation}
\label{freqEstimateMin}
\int_0^{T/4} J(t^\prime) \rmd t^\prime \simeq m\pi
\quad \textrm{and} \quad
\omega \simeq \frac{J_\text{avg}}{m}, \quad m=1,2,\ldots.
\end{equation}
Figure~\ref{freqEstimate}(a) summarizes the transfer efficiencies for a range of mechanical oscillation frequencies.
The estimated frequencies for extremal transfer agree well with the obtained numerical integration, even for the situation in the plot, in which the decay rates do not fulfill the assumption underlying the analytical solution in \eqref{sitePop1} and~\eqref{sitePop2}.

In case the combined decay rates are not equal at all the sites, the symmetry of the problem is broken, and the eigenstates of the Hamiltonian, the symmetric and anti-symmetric superposition of the excitation being at sites 1 and~2, are coupled to one-another by the decay process.
Moreover, if the initial phase is not $\pi/2$, maxima of $P_\text{sink}(\omega)$ are no longer characterized by a perfect locking of the excitation at site 2 for all times, but only during the first few relevant oscillations until the excitation has essentially left the system.

\paragraph{}
In order to quantitatively demonstrate how the presence of mechanical oscillations genuinely enhances the energy transfer, we compare the asymptotic population of the sink $P_\text{sink}(\omega)$ for a chain oscillating with frequency $\omega$ with the obtained sink population for a fixed chain in the closest and hence strongest coupled site configuration, $P_\text{sink}^\text{st}(J_\text{max})$.
For this purpose we introduce the \emph{motion-induced quantum enhancement}
\begin{equation}
\Delta(\omega)=P_\text{sink}(\omega)-P_\text{sink}^\text{st}(J_\text{max}) .
\end{equation}
We choose a set of parameters for all the following simulations that allow us to conveniently identify the effects. The chosen time scales are approximately of the same order: $J_{0}=1$, $\gamma_n=1/10$, and set $\gamma_{S}=1/2$.
Figure~\ref{freqEstimate}(a) shows that for the moving molecules the sink population $P_\text{sink}(\omega)$ surpasses the maximal static case $P_\text{sink}^\text{st}(J_\text{max})$ by several percent for the chosen parameters, and exhibits an enhancement $\Delta>0$.

\paragraph{}
We emphasize that the motion-induced enhancement beyond the efficiency of the static case at maximal coupling strength is a true quantum effect. That is, the motion-induced efficiency enhancement in the quantum mechanical case is stronger than the motion-induced enhancement in the classical case.
The coherent excitation transfer (Rabi oscillation) can transfer \emph{all} of the excitation to the second site.
A modulation of the coupling strength on a comparable time-scale may then prolong the effective time that the excitation is localized close to the sink as compared to the static chain.
In the corresponding classical case, however, the excitation transfer takes place according to F\"orster theory~\cite{Forster}, and is governed by a rate equation, e.g.\
\begin{equation}
\frac{d P_n(t)}{dt} = \sum_{m=1}^N M_{nm}(t) P_m(t),
\end{equation}
where the population of the $n$-th site $P_n(t)$ depends on the transfer rates $M_{nm}$ for a stochastic hopping between sites $n$ and $m$.
For detailed balance $M_{nm}=M_{mn}$, this leads to a diffusive propagation of the excitation through the network of sites, and in absence of dissipation to an equilibration of the site populations.
The maximal population at the last site that is connected to the sink is therefore at best $P_N(t)=1/N$.
A modulation of hopping rates $M_{nm}(t)$ does only affect the speed with which this equilibration is reached but not the maximal population of the last site.
The mechanical oscillation of the sites which modulates the hopping rates can therefore not achieve a hopping rate above the closest site configuration, where the maximal population of the last site is reached fastest, and maintained highest in presence of dissipation and decay to the sink.
The nature of diffusive transport does therefore not allow for dynamic excitation locking strategies as pictured earlier.
The corresponding quantity $\Delta_\text{cl}(\omega)=P_\text{sink,cl}(\omega)-P_\text{sink,cl}^\text{st}(J_\text{max})$ then measures the efficiency of classical diffusive excitation transport through a moving molecule structure in comparison to classical transport through a static, maximally coupled structure. 
We can therefore conclude that a value of $\Delta_\text{(cl)}(\omega)>0$ is classically forbidden.
A positive motion-induced quantum enhancement $\Delta(\omega)>0$ thus proves and quantifies a ``quantum advantage'' due to the quantum nature of the transport process.

\paragraph{}
The observed motion-induced efficiency enhancement is present for all values of amplitudes $a$ and magnitudes of the modulation of the coupling $J$.
However, the effect is less pronounced for smaller values than the ones used here.
The optimum oscillating frequency $\omega$ is then no longer characterized by a complete locking of the excitation at the second site, but rather by a modulation of the Rabi oscillation such that the excitation lingers slightly longer at the last site.
Our particular choice of $a=1/4$ amplifies the visibility of the effect, but it does not cause its presence.
Figure~\ref{freqEstimate}(b) collects the maximal enhancement $\Delta(\omega_\text{opt})$ at the optimal frequency for several values of the amplitude $a$.
Even for reasonably small amplitudes of the displacement, the effect is of the order of a percent.
The doubly logarithmic plot suggests that the effect diminishes only algebraically for amplitudes ranging from large displacements that are expected  for a driven motion, down to small position fluctuations as caused by thermal effects.

In $\alpha$-helices the equilibrium distance between the amide-I oscillators is $d_0\simeq4.5$\,\AA{} whereas their relative displacement reaches values of $d_0a\sim0.1$\,\AA~\cite{Scott92,CruzeiroReview}, which amounts to an $a$ of one order of magnitude smaller.
Nevertheless, the coupling $J$ exhibits a variation of $\sim7\%$.
For smaller amplitudes of the mechanical oscillation than the ones chosen to obtain figure~\ref{freqEstimate}, the extrema persist but are less pronounced.
It is therefore conceivable that, at the right frequency of motion, even in biological or chemical systems where oscillation amplitudes are small, the quantum-coherent nature of this transport process on a moving structure yields efficiency improvements beyond what is classically achievable.
Please also note that the total efficiency gain with respect to the equilibrium situation is much larger.

\begin{figure}
\centering
\includegraphics[width=0.8\textwidth]{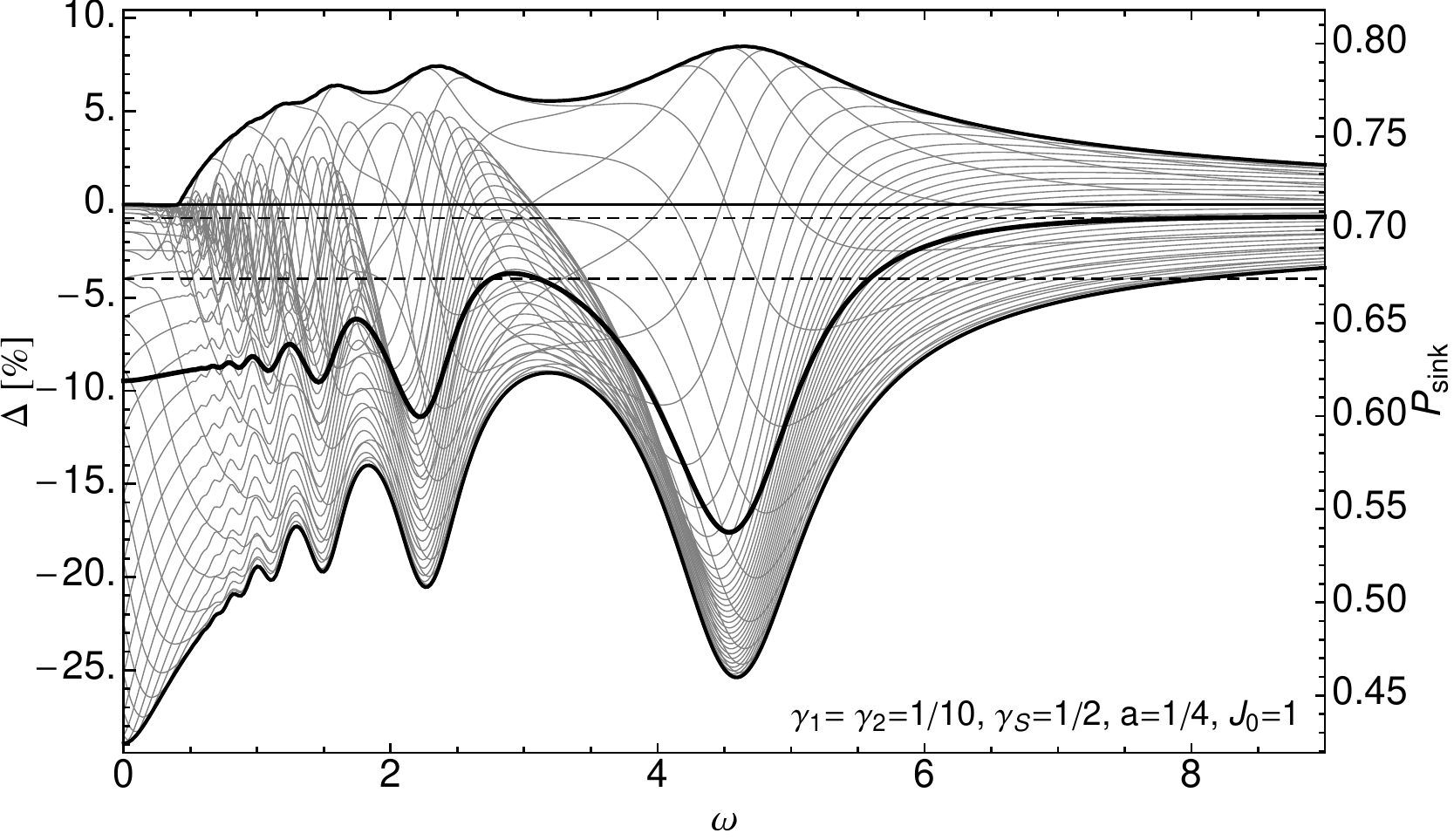}
\caption{Asymptotic sink population $P_\text{sink}(\omega)$ for different initial phases $\phi$ of the mechanical oscillation.
Thin grey curves correspond to 50 initial phases uniformly spaced in $\phi\in[0, 2\pi]$. Thick black curves indicate the enveloping curves (outer curves), and the phase averaged asymptotic sink population (inner curve) of 200 uniformly sampled phases. Horizontal lines again show the sink population for static cases with $J_\text{max}$, $J_\text{avg}$, and $J_0$ (from top to bottom).
For asymptotically large $\omega$, all curves approach the sink population of the static case with $J_\text{avg}$.
}
\label{2sites_phases}
\end{figure}

So far, the discussion was limited to the motion of the molecules starting with the initial phase~$\phi=\pi/2$, i.e.\ with the molecule in the closest and hence maximally coupled configuration.
This is justified -- as explained earlier and shown in figure~\ref{chainmodel}~(right) -- when the motion is triggered by the excitation of the complex that changes the nuclear equilibrium position to greater distances than in the ground state.
For motion of mostly thermal origin, such a coordination is generally not the case, and the initial phase would be random.
In particular, one cannot expect an enhancement if the molecules are initially in a distant configuration and move away from each other.
We therefore expect that, in the absence of synchronization between excitation transfer and motion, the molecules will not be able to efficiently drive the transfer.

In order to test whether or not an enhancement of the excitation transfer above any comparable static case persists for an arbitrary initial phase of the mechanical oscillation, we statistically sample uniformly over initial phases.
In figure~\ref{2sites_phases}, each thin gray curve represents the sink population and enhancement for a given initial phase of the oscillation, of which the figure~\ref{freqEstimate} represents an instance.
The outer black curves are the envelopes of the data, and represent the maximal and minimal sink population that can be obtained for the chosen parameters at a given oscillation frequency~$\omega$.
The inner black curve shows the average sink population over all initial phases.
We can observe that the average sink population for a random initial phase lies below the static case with coupling $J_0$ for low frequencies, and approaches the efficiency of the static case with the time-averaged coupling strength $J_\text{avg}$ from below for large frequencies.
An enhancement of the transfer efficiency over the static case with $J_0$ can on average still exist simply due to the fact that the average coupling strength is stronger than $J_0$.
Furthermore, it becomes evident that an enhancement $\Delta>0$ can only be reached in a window of frequencies.
For very slow motion, i.e.\ for small~$\omega$, the excitation propagates on a quasi-static structure, and a certain threshold of the underlying motion has to be passed before $\Delta>0$ appears.
For very fast motion, $\omega\gg J_0$ the transport efficiencies converge towards the static configuration with the time-averaged coupling strength and thus $\Delta<0$ (not shown in figure~\ref{2sites_phases}).
We can thus conclude that for a given oscillation frequency $\omega\simeq O(J_0)$, a large enhancement over the static situation can be obtained, but strongly depends on the initial phase.
Therefore, it is crucial that the transport process can be initialized with a specific range of phases of the site oscillation rather than a uniformly distributed random phase.
At the same time, however, if one has an experimental way of setting or influencing the initial phase, the strong phase-dependence might be exploited to switch or to observe this effect experimentally.

\paragraph*{}
In the following sections, we consider longer chains and perform a deeper analysis of the behavior of $\Delta(\omega)$.
We anticipate that the main features identified in the dimer model are qualitatively present also in larger systems, and that the quick-transfer-and-locking strategy can be employed to shuttle the excitation along a longer structure.
In particular, we expect that $\Delta(\omega)$ exhibits an analogous oscillatory behavior that extends to positive values for a specific range of frequencies.
The identification of such optimal frequencies could in principle lead to a better control and to an enhancement of the efficiency of energy transfer in artificially designed systems via the modulation of the oscillation frequency.

\section{Multi-molecule oscillations}
\label{standing oscillations}

In this section, we consider longer linear chains as a first step towards a more suitable representation of nature-inspired systems, such as the amide-I oscillators along an $\alpha$-helix or the chromophores in a light harvesting complex.
The analysis performed in the dimer case is now extended to linear chains of $N$ molecules.
The underlying intuition is that the molecules carrying the excitation are embedded in a larger complex, e.g.\ a protein, but again are assumed to follow a concerted motion.
The particular environmental and boundary conditions change from instance to instance.
For this reason and for simplicity, we model the linear chain as a succession of molecules having the same mass and attached via a spring to their respective neighbors, as e.g.\ inspired by the amide units in an $\alpha$-helix.
For the first and last site we consider two possible situations:
both the ends are (i) completely free or (ii) both are in contact with two ``walls'' via two additional springs.
In the first case, the chain would represent an $\alpha$-helix structure without a narrow confinement in the longitudinal direction, while in the latter case the form of the surrounding protein determines an effective space in which the helix is confined.
For both situations, we study collective normal mode oscillations of low frequency as a relevant example of naturally occurring oscillations.

The transport efficiencies along the chain that are obtained under these dynamical conditions are compared to the ones in the static case, in which all the molecules are at the equilibrium distance.
We thus obtain the gain that is simply due to the presence of motion.
Alternatively, we compare to a configuration where the sites are maximally coupled, which yields~$\Delta$.
In the case of uniformly coupled, static chains of more than three sites, the excitation is no longer completely transferred to the last site with probability one, from where it may be trapped into the sink~\cite{Datta}.
Since, in addition, with increasing chain length the excitation is also more likely to decay into the environment, we expect lower absolute efficiencies than in the dimer toy model.
On the other hand, there is the possibility that the motion-induced enhancement might add up over several transfer step along the chain.

\begin{figure}
\centering
\includegraphics[width=0.45\textwidth]{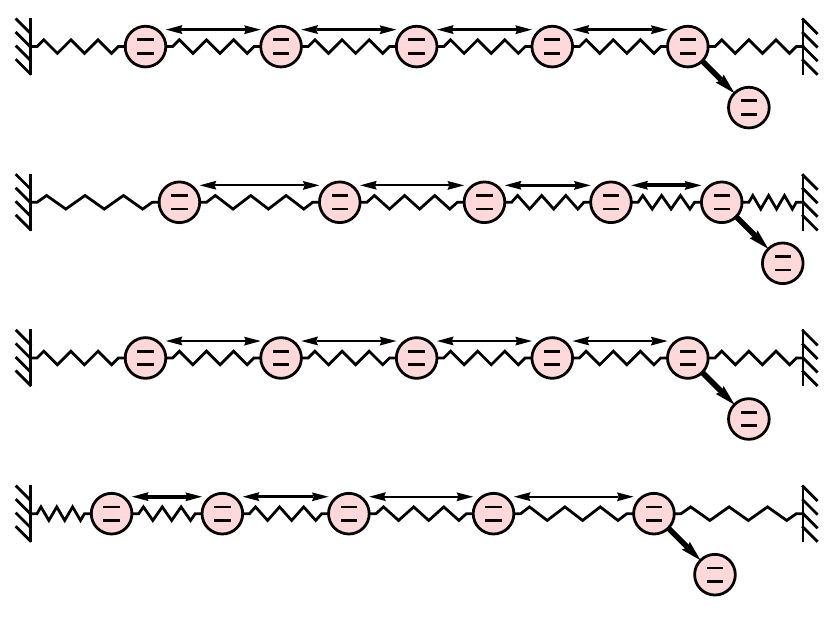}\hfill
\includegraphics[width=0.45\textwidth]{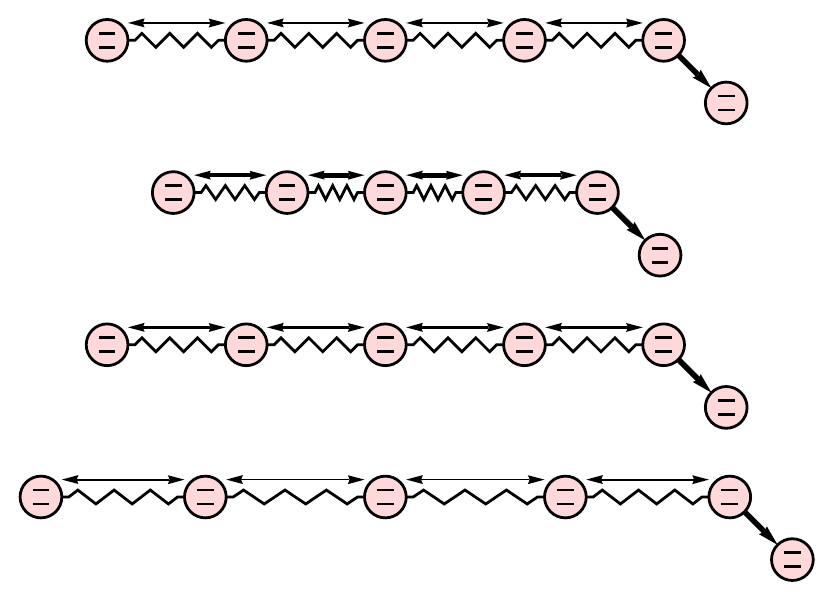}
\caption{Left: Sloshing mode of oscillation in which half of the system of two-level sites contracts and half expands alternately. Right:~Breathing mode in which the entire system contracts and expands periodically. The individual sites are coherently coupled, and the sink is connected via irreversible decay to the last site.}
\label{modes}
\end{figure}

Before comparing the asymptotic amounts of population in the sink, we briefly introduce the model that we use to derive the proper motion of the linear chains.
For simplicity, we assume that the forces between two neighboring sites are proportional to their relative displacements from the equilibrium positions:
this approximation, known as the harmonic approximation, holds as long as the displacements are relatively small.
To help in visualizing the system, one may think of the molecules forming the unidimensional lattice as connected by elastic springs (see figure~\ref{modes}).
In the chain of coupled oscillators, the internal force exerted on the $n$-th site is:
\begin{equation}
F_{n}=M\frac{d^{2}u_{n}}{dt^{2}}=K(u_{n-1}-u_{n})+K(u_{n+1}-u_{n}),
\end{equation}
where $K$ is the spring constant (intermolecular force), $M$ is the particle mass, and $u_{n}$ the displacement of the $n$-th site from its equilibrium position.
For the first and last site, we alternatively have only one spring or fix them with additional springs to the confining boundaries.
At this point we neglect the influence of random thermal forces on the particles.
An additional noise term in the position and thus in the coupling strength, as long as it is sufficiently small, will only slightly affect the effective Rabi frequency as compared to the case without noise, since the Rabi frequency it is effectively obtained by integrating over the coupling strength.
The particular shape of $J_n(t)$, and fluctuations of it will therefore not greatly influence the transport efficiency as long as a quick-transfer-and-locking strategy can be achieved.

A chain of $N$ coupled oscillators has a generic solution of the above equations of motion that can be expressed as a linear combination of normal modes.
The normal modes are independent, collective modes of oscillation in which all the sites move with the same periodicity and do not cause the excitation of other oscillatory motions.
Each of these modes is characterized by a frequency and a relative phase between the individual sites.
For the confined chain, the normal mode frequencies and the displacement of the $n$-th site are (see e.g.~\cite{Rosenstock})
\begin{eqnarray}
\label{freqConfined}
\omega_{q} &=2\omega_{0}\sin\left[\frac{q\pi}{2(N+1)}\right],
\qquad \text{and} \\
\label{displConfined}
u_{n}(t) &=\sum_{q=1}^N \frac{a_q d_0}{A_q} \sin\left(\frac{q\pi}{N+1}n\right) \sin(\omega_{q}t+\phi_q),
\end{eqnarray}
respectively, where $\omega_{0}=\sqrt{K/M}$, $A_q=\sin[q\pi/(N+1)]$ is a normalization constant, and $d_0$ is the equilibrium distance between neighboring sites.
For the chain with open ends the expressions are
\begin{eqnarray}
\label{freqOpen}
\omega_{q} &=2\omega_{0}\sin\left(\frac{q\pi}{2N}\right),
\qquad \text{and} \\
\label{displOpen}
u_{n}(t) &=\sum_{q=1}^N \frac{a_q d_0}{B_q} \cos\left[\frac{q\pi}{N}\left(n-\frac{1}{2}\right)\right] \sin(\omega_{q}t+\phi_q),
\end{eqnarray}
with $B_q=\cos[q\pi/(2N)]$.
Figure~\ref{modes} illustrates the normal mode with the lowest frequency for the confined chain, the sloshing mode, and for the open chain, the stretching or breathing mode.

\begin{figure}
\centering
\includegraphics[width=0.7\textwidth]{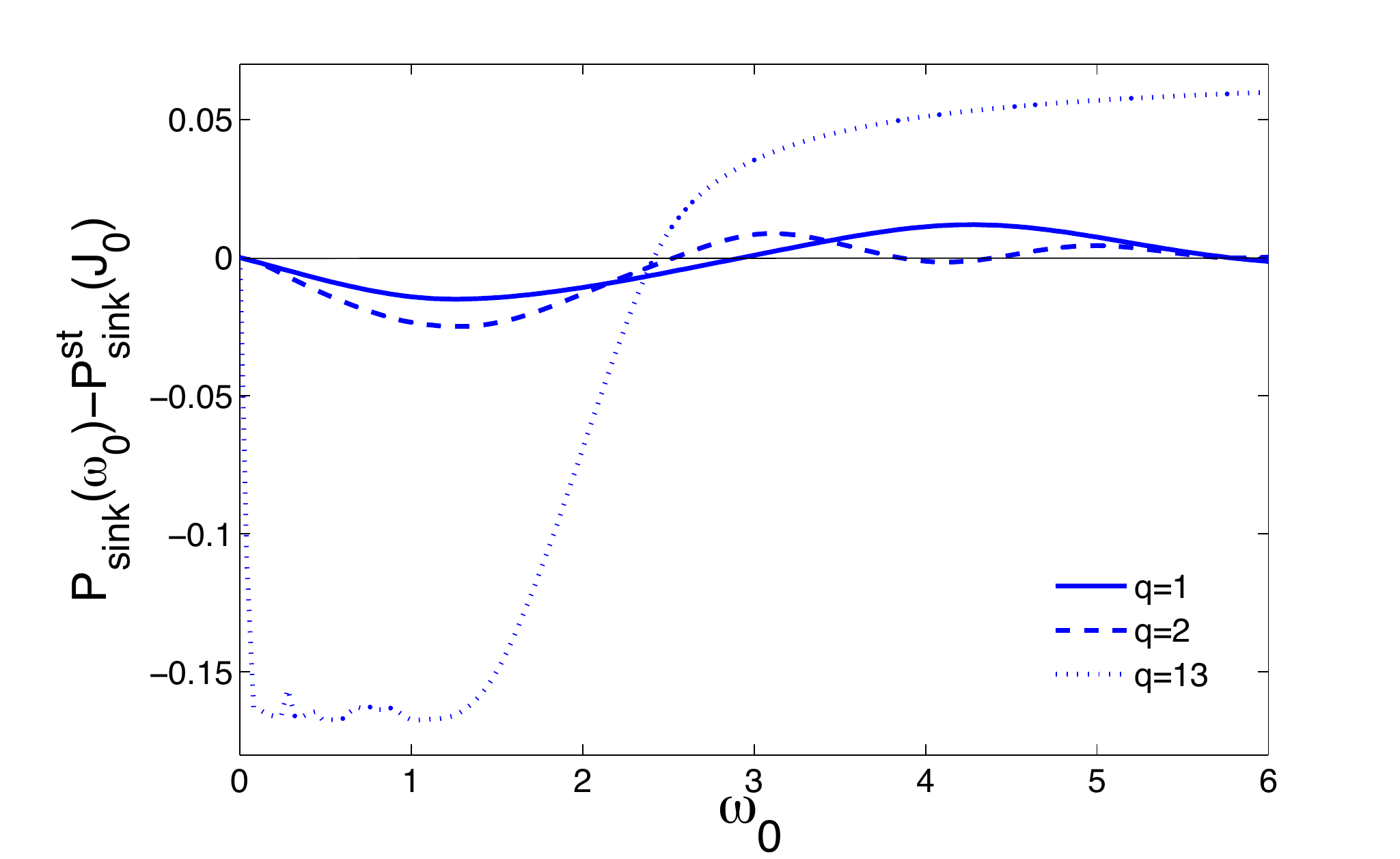}
\caption{Motion induced energy transfer gain over the uniformly coupled chain for different normal modes of $N=13$ confined sites. Oscillation amplitude constant is $a_q=1/24$ and $\phi_q=\pi$.
Only the modes $q=1, 2$, and 13 are shown. $P_\text{sink}^\text{st}(J_0)\approx 0.17$, and the flat part of $q=13$ for low $\omega_0$ amounts to a vanishing absolute sink population.}
\label{Nsites_QEnhancement}
\end{figure}

\paragraph*{}
For a more detailed study, we first analyze the effect of oscillations for individual normal modes of the confined chain.
This approach is consistent with the fact that slow coherent motion of the underlying structure is indeed observed in proteins where transport phenomena occur~\cite{Vos93,Vos00}.
We mimic this slow coherent motion by the low frequency collective modes of the chain.
The sloshing mode, the lowest lying of the normal modes, represents a contraction of one half of the system, and a simultaneous expansion of the other half. For higher modes, the system is divided into more parts, each of which can be considered to be in a local sloshing mode.
Figure~\ref{Nsites_QEnhancement} shows the increase of the asymptotic sink population over the static case for selected normal modes of a confined chain of coupled oscillators, when varying~$\omega_0$, i.e.\ spring constant and/or mass of the molecules.
The structural features that we found for the simple dimer case persist, and exhibit frequencies with increased and suppressed transfer efficiency.

For low frequencies, a suppression in the transport efficiency dominates the behavior.
In this frequency regime, the mechanical oscillation is almost static compared to the timescale of the coherent excitation transfer.
The excitation transfer thus happens on a chain with disordered couplings, which lead to localization of the excitation at the beginning of the chain, and hence to a decreased transfer efficiency.

For higher frequencies $\omega_0$, we obtain efficiency maxima that surpass the transport efficiency of the static, ordered chain.
As in the dimer model, these frequencies are in resonance with the wave-like propagation of the excitation through the chain.

For asymptotically large $\omega_0$ all curves converge to the efficiency of the static case with the time-averaged coupling strength between nearest neighbors (not shown in figure~\ref{Nsites_QEnhancement}). Since we fix $a_q=1/24$ for all modes, and the relative distance between two sites is also determined by a position- and mode-dependent prefactor in \eqref{displConfined} and~\eqref{displOpen}, the time-averaged coupling strengths differ between the normal modes, and between different sites of a single mode.

\begin{figure}
\centering
\begin{tabular}{rr}
\includegraphics[width=.5\textwidth]{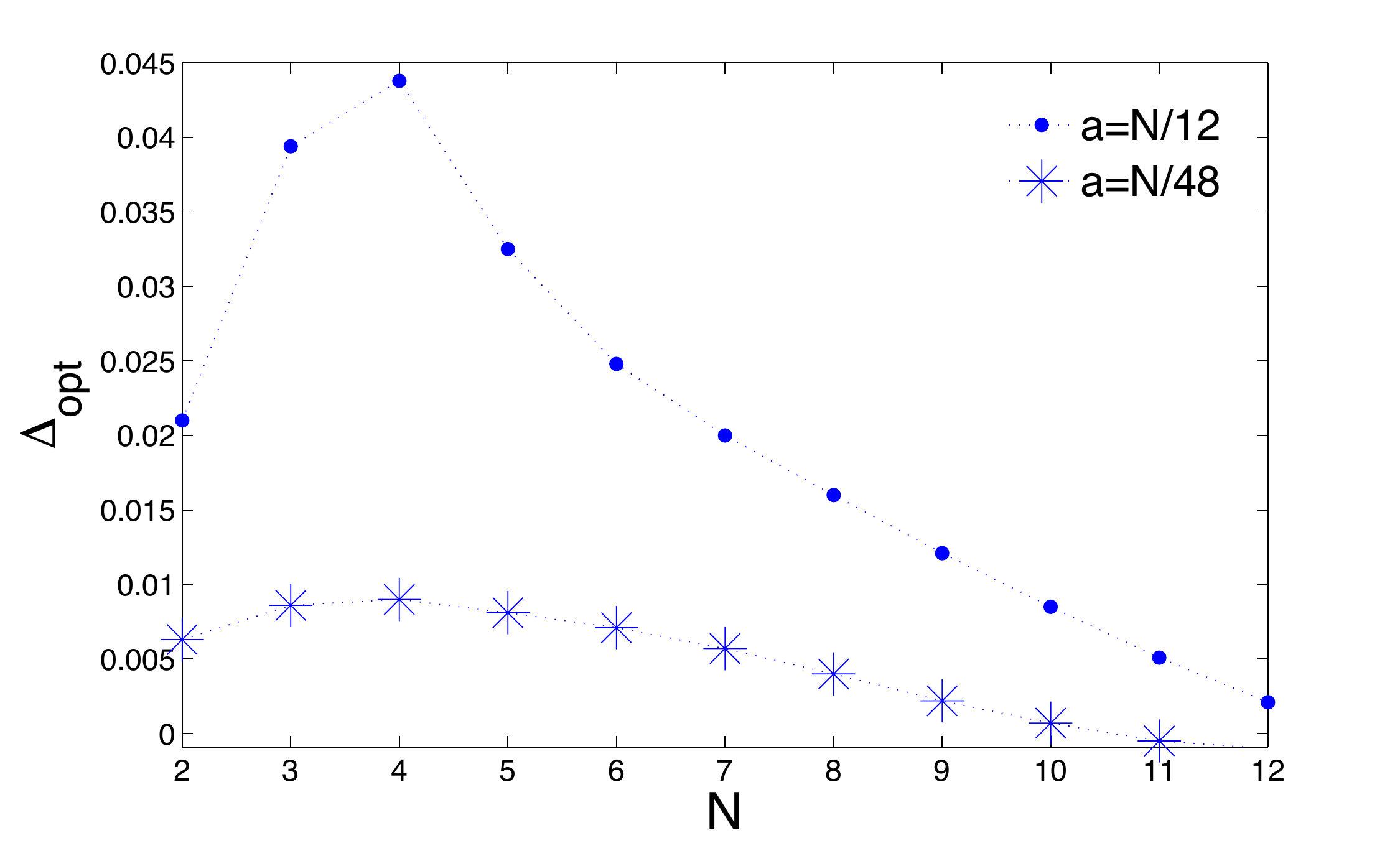} &
\includegraphics[width=.5\textwidth]{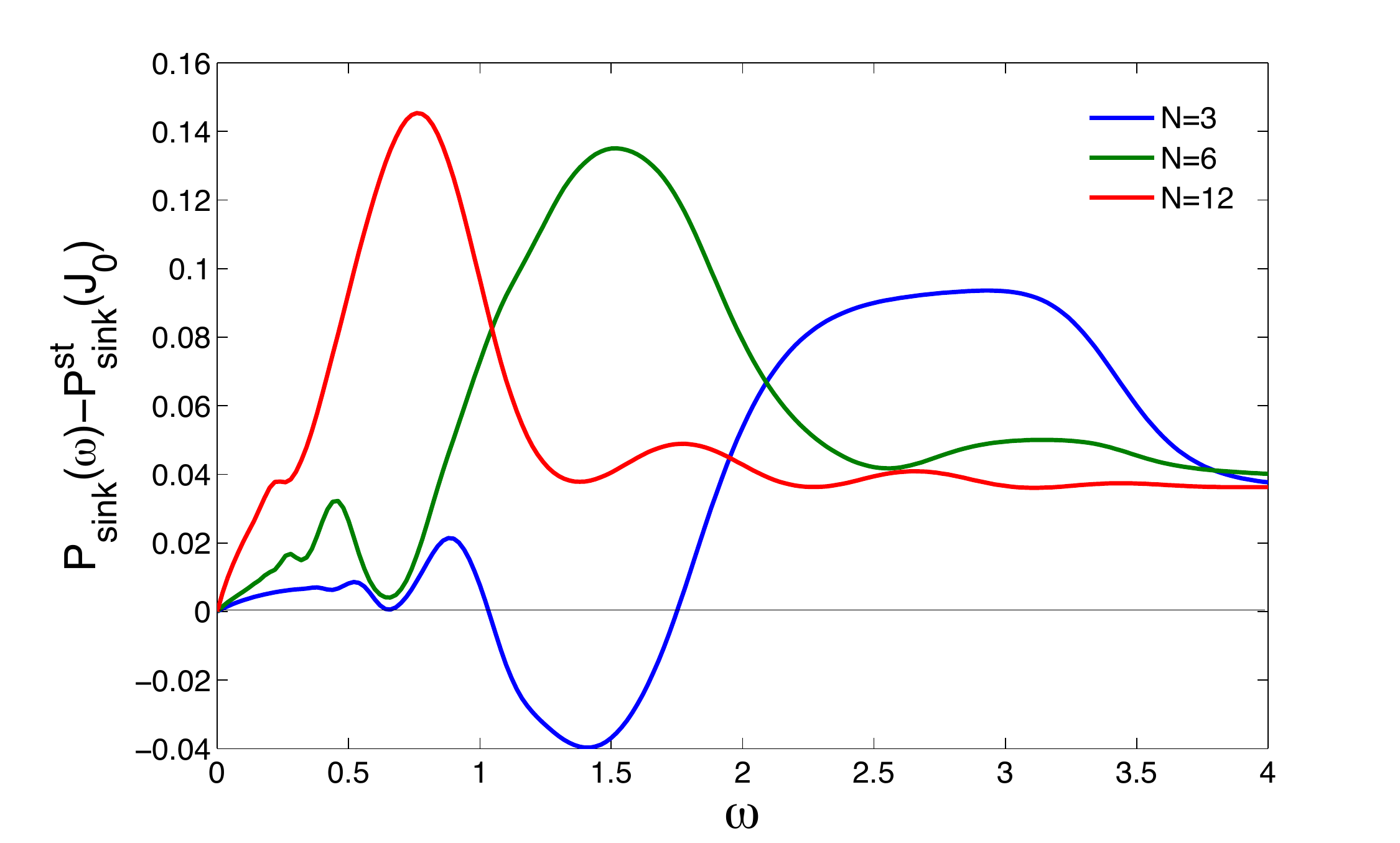} \\
\includegraphics[width=.5\textwidth]{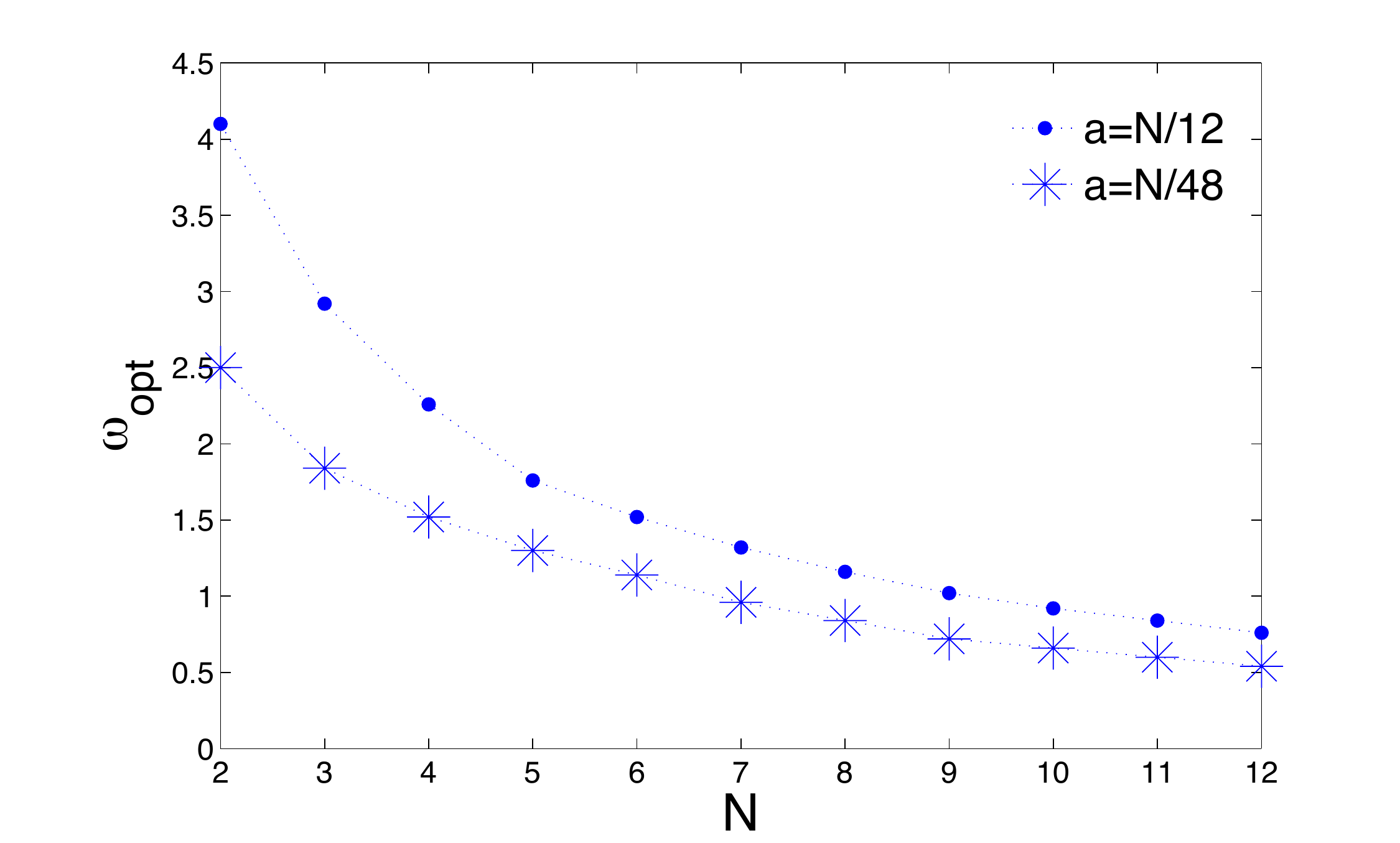} &
\includegraphics[width=.5\textwidth]{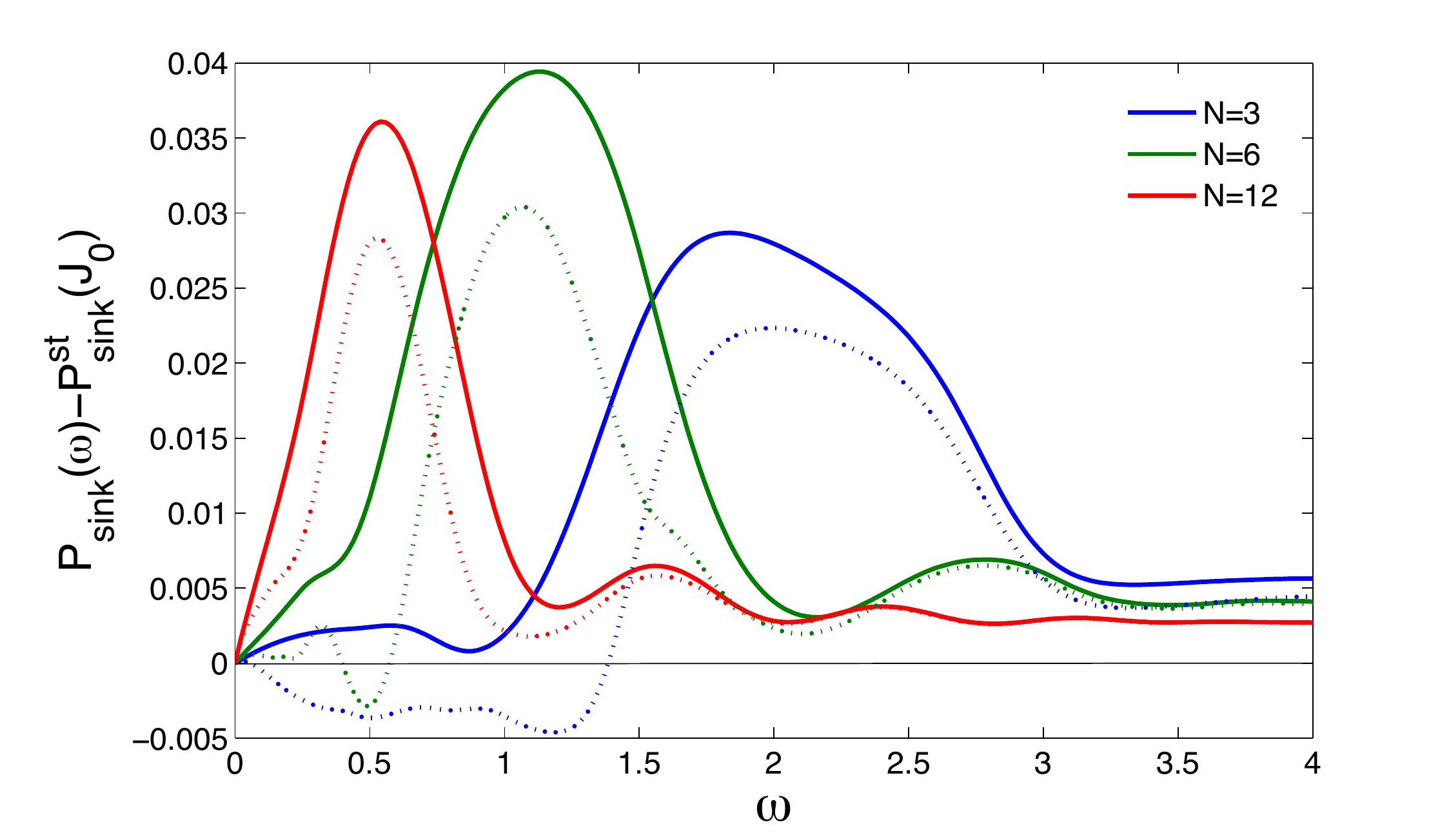}
\end{tabular}
\caption{Left: Maximum energy transfer enhancement over the maximally coupled static chain $\Delta_\text{opt}$ (top) and the corresponding optimal oscillation frequency $\omega_\text{opt}$ (bottom) for the breathing mode for different chain lengths $N$ and oscillation amplitude constants $a\equiv a_1$.
Right:~Frequency dependence ($\omega\equiv\omega_1$) of the energy transfer enhancement as compared to the chain at rest, i.e.\ for~$J_0$, with amplitude constants $a_1=N/12$ (top), $a_1=N/48$ (bottom), and $\phi_1=0$.
The dotted lines include a detuning of the site energies due to the motion with coupling strength $\chi=10$.}
\label{NSites_optimalEnhancement}
\end{figure}

\paragraph*{}
In order to elucidate the dependence of the transport efficiency on the parameters that define the system, we investigate different amplitudes and chain lengths for a single normal mode.
Since the sloshing mode has no configuration where all the sites are closest and hence maximally coupled (half of the chain expands whereas the other half is compressed), we choose the breathing mode of the unconfined chain, where there exists such a configuration.
For the breathing mode, figure~\ref{NSites_optimalEnhancement} collects the data of the maximum enhancement of the transfer efficiency over the maximally coupled static case, i.e.\ $\Delta_\text{opt}=\max_{\omega_1}P_\text{sink}(\omega_1)-P_\text{sink}^\text{st}(J_{n,\text{max}})$, and the corresponding optimal oscillation frequency $\omega_\text{opt}$.
The amplitude constant $a\equiv a_1$ is chosen such that the maximal extension of the chain increases linearly with $N$, and the relative change in length of the entire chain during the mechanical oscillation, $\Delta L/L$, remains constant when changing~$N$.
In particular, the case of $a=N/48$ provides an instance where the relative distance between neighboring molecules is of the order of what is also found in vibrating $\alpha$-helices at room temperature~\cite{Pleiss91}.

\begin{figure}
\centering
\includegraphics[height=5cm]{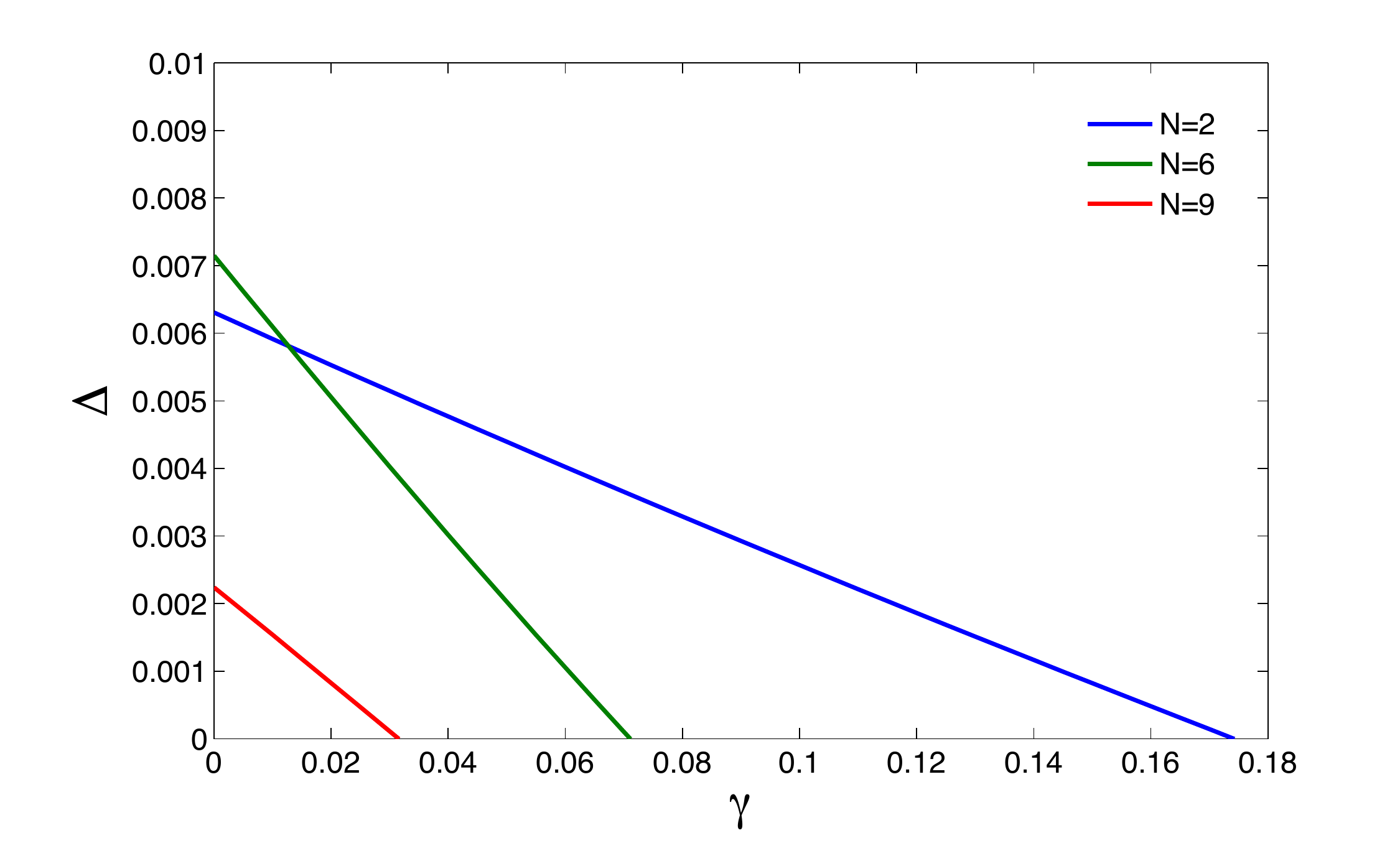}%
\includegraphics[height=5cm]{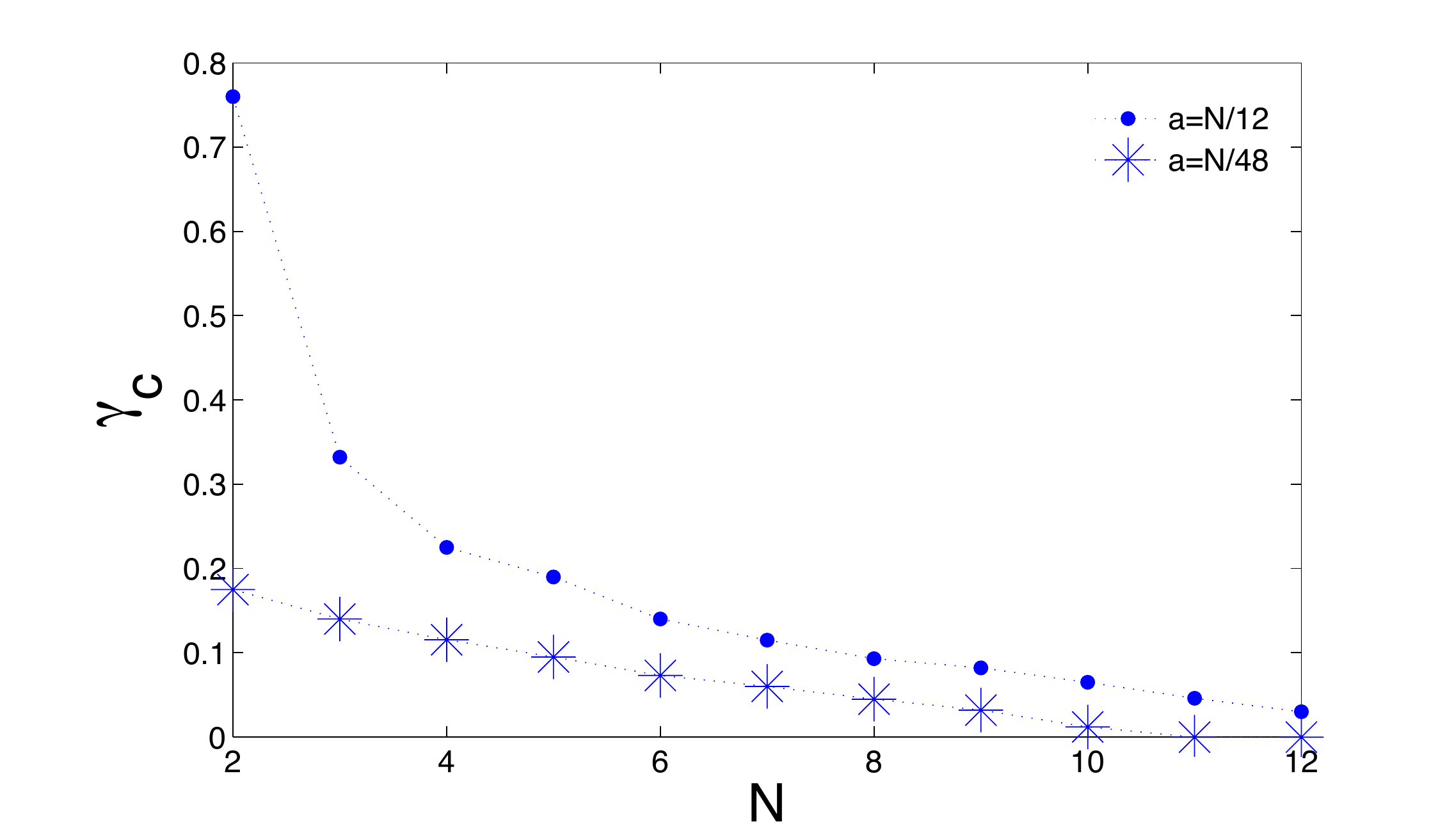}
\caption{Left: Decay of the motion-induced transport enhancement $\Delta$ for the breathing mode as a function of the dephasing rate for different chain lengths. Amplitude constant is $a\equiv a_1=N/48$, $\phi_1=0$, at a fixed frequency~$\omega_0$ that is optimal at $\gamma=0$.
Right:~Critical dephasing rate $\gamma_{c}$ dependent on the chain length $N$.}
\label{NSites_dephasing1}
\end{figure}

As for the dimer molecule, a positive $\Delta_\text{opt}$ cannot be achieved in the analogous classical description, and likewise here, for the quantum-coherent transport, positive values exist over a large range of chain lengths.
We observe that, in general, $\Delta_\text{opt}$ decreases with increasing $N$.
This is essentially due to the larger distance that the excitation needs to travel, and hence a proportionally increasing probability that the excitation is dissipated.
We attribute the increase for short chain lengths to the accumulation of the efficiency gain, which is observed for two sites, over several sites.
The decrease of $\omega_\text{opt}$ with increasing chain length qualitatively follows the inverse of the time at which the excitation is (partially) transferred to the last site in a homogeneously coupled chain, i.e.\ with increasing chain length, the time also increases at which the first maximum of the last site's population appears~\cite{Datta}.
We can apply a similarly intuitive argument to motivate the dependency of $\omega_\text{opt}$ on the oscillation amplitude factor~$a$.
When the amplitude factor $a$ of the oscillation is large, the inter-site coupling strength is subject to a large variation and therefore, when the molecules are closest to each other, the excitation propagates very fast such that a quick modulation of coupling strength is necessary to prevent the immediate reversion of the population.

We expect that this non-classical enhancement induced by oscillations can be observed not only for the chains with uniform site energy (i.e.\ with $\varepsilon_{n}=\varepsilon \ \forall n$), but also for systems with local energy disorder.
There, the varying coupling strength may effectively lift the disorder, since the localization depends on the ratio of detuning and coupling strength.

It is also worth mentioning that in the regime where $J_0\ll\gamma_{S}$, i.e.\ for a large decay rate into the sink, a quantum Zeno-type effect slows down, paradoxically, the absorption into the sink by projecting the system with high probability into a state of an unpopulated last site.
Therefore, in this regime, variations of the inter-site coupling strength due to oscillations have no pronounced effect on energy transfer efficiency.
On the other hand, in the regime of a weak sink rate and weak dissipation, the excitation undergoes many cycles before it is finally absorbed, thus amplifying the effect of the oscillations.

In order to test, how a local detuning of the (at equilibrium position) uniform site energies affects the observed transport efficiency enhancement, we add an exciton-vibration coupling term,
\begin{equation}
\label{exphCoupling}
H_\text{ex-vib} = \chi \Big(u_{n+1}(t)-u_n(t)\Big) \proj{n}{n},
\end{equation}
which also appears in $\alpha$-helices~\cite{Scott92}, to the Hamiltonian~\eqref{Hamiltonian}.
The coupling strength $\chi=10$ is chosen such that for $a=1/4$ in the dimer, the energies are of the same order as the exciton coupling between the sites.
Figure~\ref{NSites_optimalEnhancement} (bottom,right) compares the transfer efficiencies with (dotted lines) and without this detuning (solid).
Although the motion-induced efficiency gain suffers due to the detuning of the site energies and the thereby caused localization, the effect persists.

\begin{figure}
\centering
\includegraphics[height=5cm]{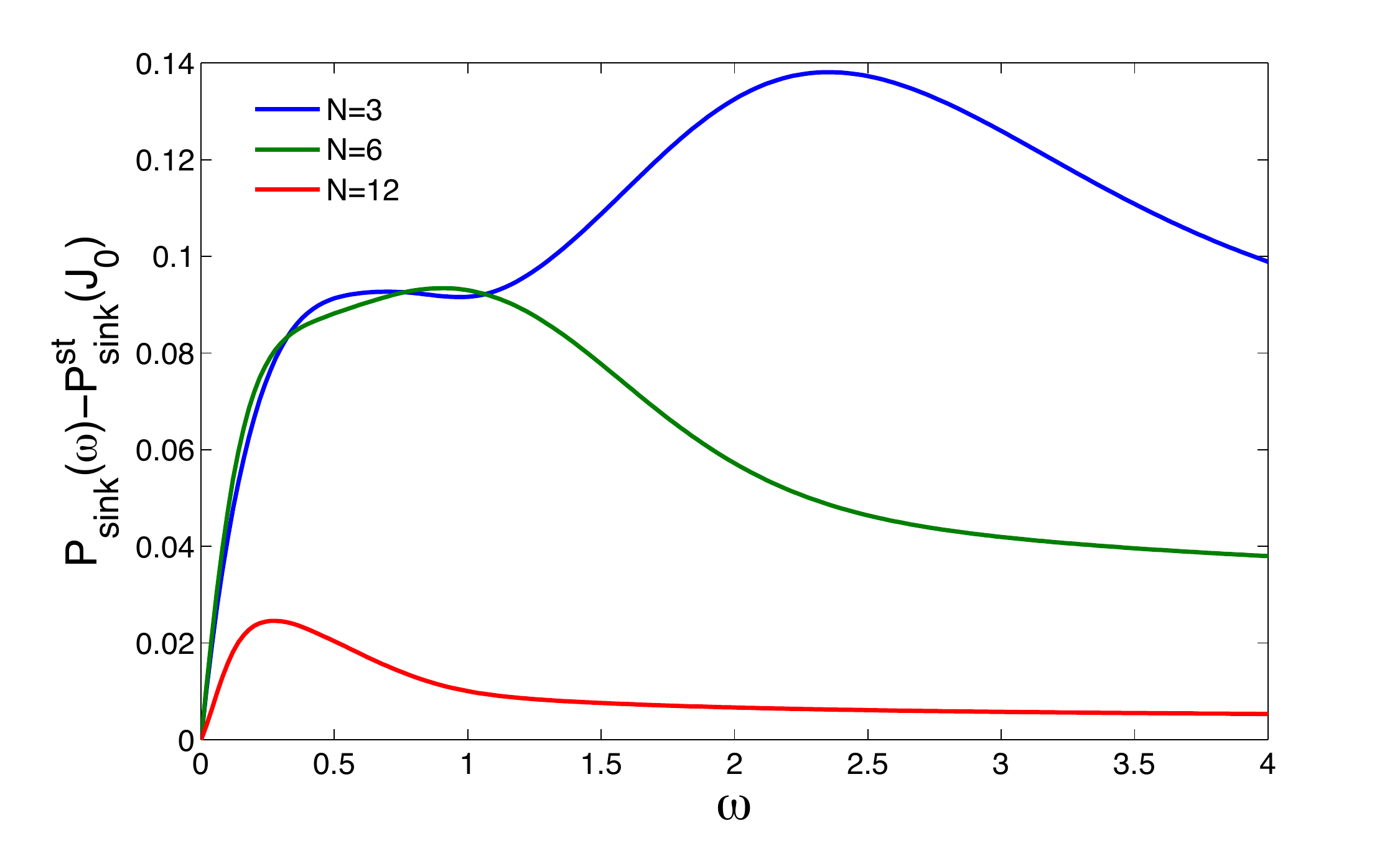}%
\includegraphics[height=5cm]{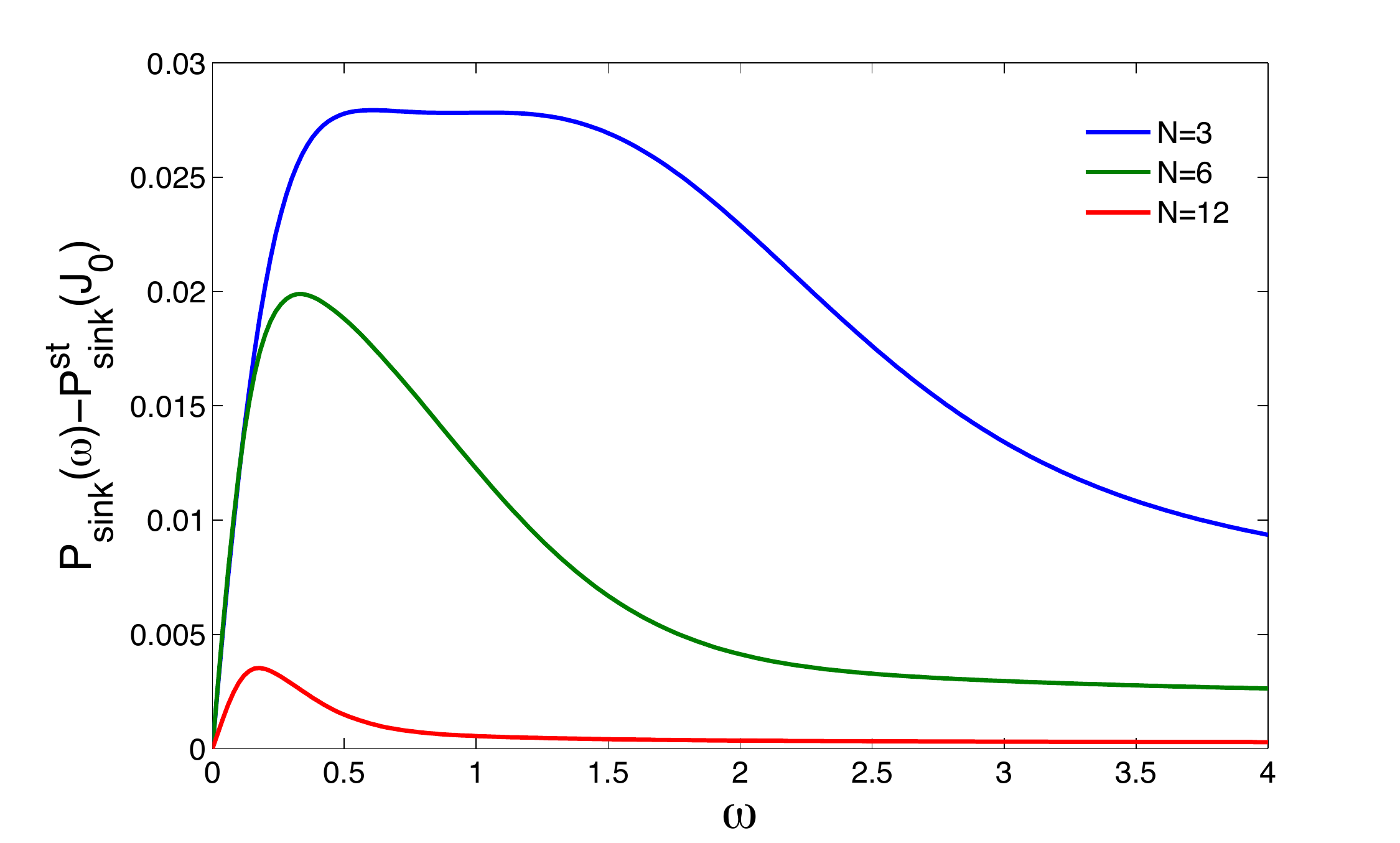}
\caption{Enhancement of energy transfer efficiency of the oscillating chain (breathing mode) as compared to the transfer efficiency of the resting chain coupled with $J_0$ as both suffer strong dephasing at rate $\gamma=1$ with $a_1=N/12$ (left) and $a_1=N/48$ (right), and $\phi_1=0$.}
\label{NSites_dephasing2}
\end{figure}

The claim that the observed enhancement is indeed a quantum feature is tested by adding a dephasing environment, which in general leads to the loss of quantum coherence.
This is modeled by an additional term in the master equation~\eqref{mastereq}:
\begin{equation}
\label{dephasing}
L_\text{deph}\rho= \gamma \sum_{n=1}^N \left(2\sigma^+_n\sigma^-_n\rho\sigma^+_n\sigma^-_n -\{\sigma^+_n\sigma^-_n,\rho\}\right)
\end{equation}
Since we expect quantum coherence to be a key requirement on which this effect relies, adding a dephasing environment should lead to a smaller enhancement.
Indeed, the simulations in figure~\ref{NSites_dephasing1} show that the enhancement over the static case with $J_\text{max}$ disappears at a specific critical dephasing rate $\gamma_{c}$. However, an enhancement of the transport efficiency over the static case with $J_0$ still exists, even for large dephasing rates (see figure~\ref{NSites_dephasing2}).
As expected, the critical dephasing rate becomes smaller as the system size increases due to the longer time over which the excitation is exposed to the coherence-destroying action of the environment.
Please note, that in \cite{Plenio08,Caruso09} an enhancement of the transport in the presence of dephasing is observed when adding dephasing to a static, disordered transport network.
In contrast, the enhancement observed here comes from the concerted motion of the molecules, and it is an enhancement with respect to the static case with the same dephasing.
This enhancement exists whether or not dephasing is present.
%
%

\section{Guided excitation transfer}
\label{guided transfer}

In the previous sections, we observed that motion and in particular oscillations can nontrivially interact with the quantum dynamics of the excitation such as to drive and enhance the energy transfer.
We now extend the previous cases to an externally driven scenario.
As a simple example, we investigate the sweeping of excited population guided by a Gaussian pulse that modifies the displacement and hence the interaction between the molecules as it travels along the chain.
A toy model could be implemented in a simulation with trapped ions by shining a laser with a suitable spatial intensity profile onto the chain, for example.
In a molecular context, a conceivable mean could be to attach the molecular chain to a nano-mechanical oscillator.

We model the pulse such that the distance between molecules $n$ and $n+1$, $d_n(t)=d_0-u(t)_{n,n+1}$, is changed from their equilibrium distance by
\begin{equation}
\label{guidedPulse}
u(t)_{n,n+1}=A\,d_0 \exp\left(-\frac{[(n-1) d_{0}-vt]^2}{2\sigma^2}\right),
\end{equation}
where $\sigma$ is the pulse width, and $v$ its velocity.
The displacement due to the pulse remains Gaussian as the pulse uniformly propagates along the chain with its center at the position of site $n$, i.e.~$nd_{0}=vt+d_0$. The effect of this pulse is to concentrate the molecules in a certain region of the chain and thereby couple them more strongly, whereas the molecule density and coupling remains unaffected outside the pulse.
Again, we would like to stress that there exist alternative handles to modulate the coupling strength, for example, a suitable change in the direction of the molecules' dipole moments.

\begin{figure}
\centering
\includegraphics[width=0.5\textwidth]{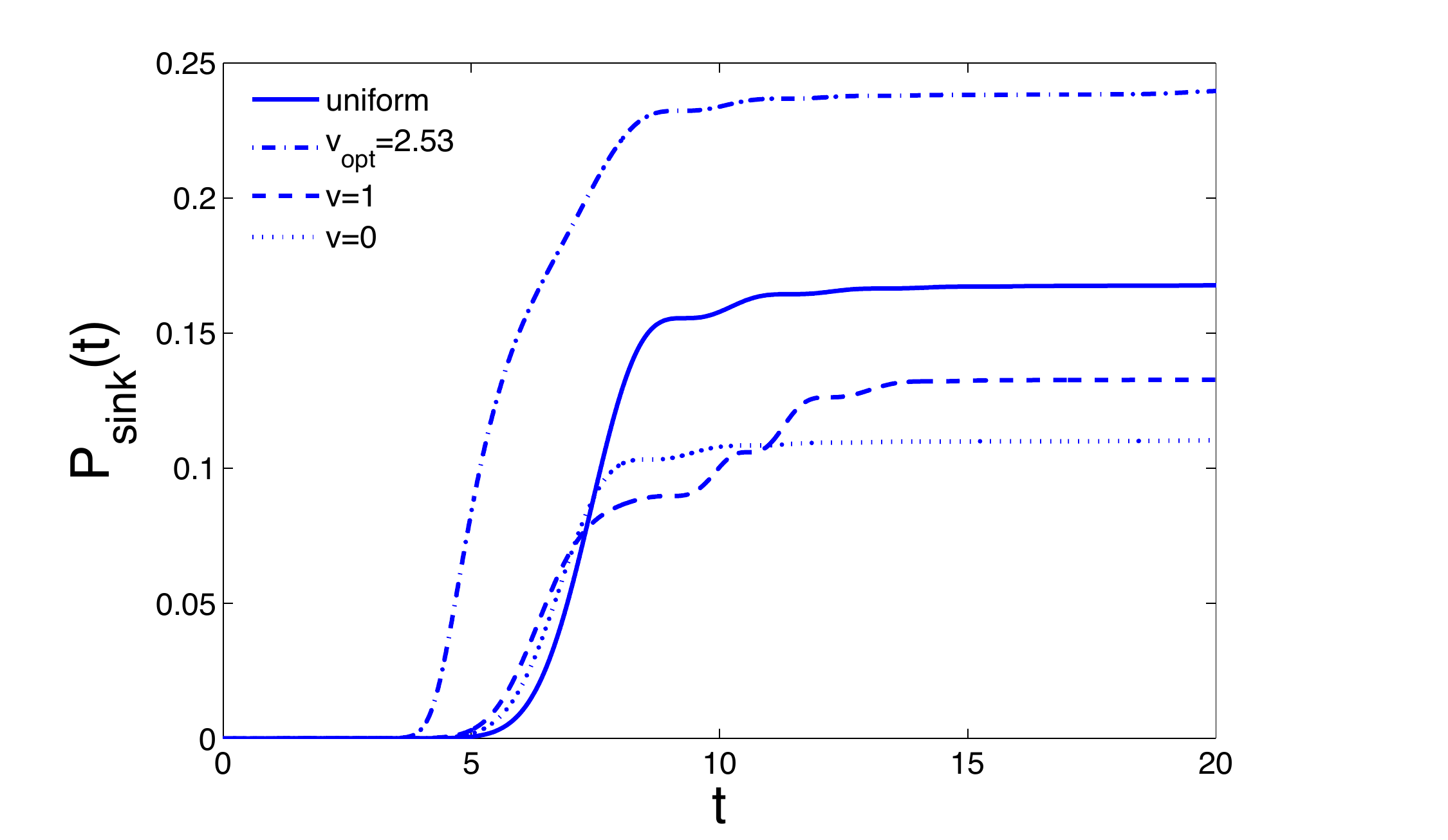}%
\includegraphics[width=0.5\textwidth]{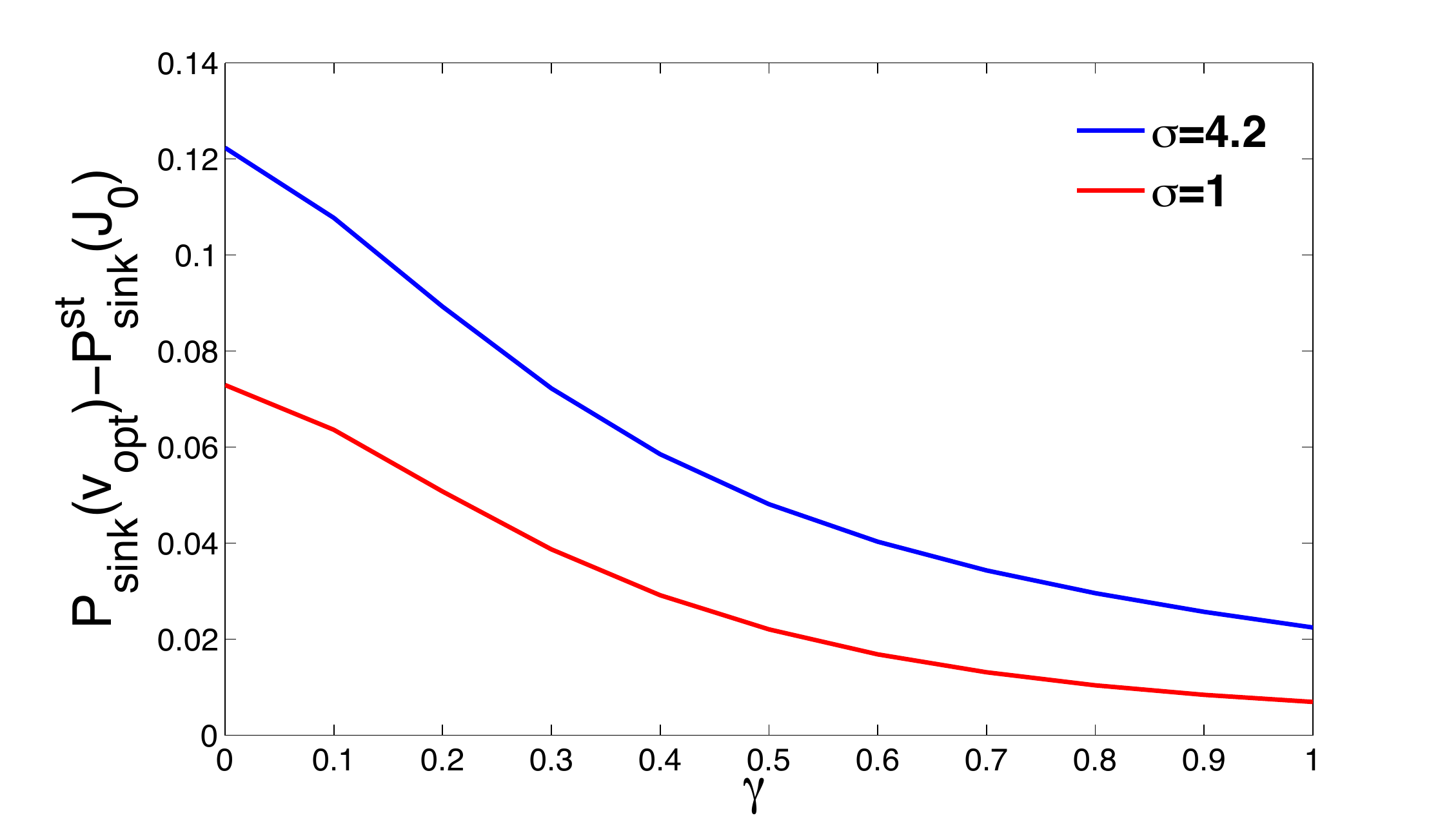}
\caption{Left: Sink population for a Gaussian wave packet sweeping along the coupled chain of $N=13$ sites. The solid line marks the reference case without pulse, dashed/dotted lines indicate data for pulses of different speed, but same width $\sigma=1$, and strength $A=1/6$.
Right:~Transport efficiency gain due to the pulse when dephasing at rate $\gamma$ is taken into account. Curves are obtained for the optimal speed for each~$\gamma$.}
\label{popGuided}
\end{figure}

We start the pulse centered around site $n=1$ at time $t=0$, and first observe the moving Gaussian pulse according to~\eqref{guidedPulse} for different values of its speed~$v$.
All lengths are measured in units of $d_0$, such that we effectively set $d_0=1$.
In figure~\ref{popGuided}~(left), we monitor the sink population as a function of time for different values of the pulse speed.
The continuous line shows the reference case without a pulse, i.e.\ for the uniformly coupled chain with sites equally spaced at distance~$d_0$.
Dashed and dotted lines indicate solutions where a pulse is present.
Clearly, there is an effective interplay between the moving Gaussian pulse and the dynamics of excitation transfer.

When the speed of the Gaussian packet is zero, which means a stationary packet centered around the first site, it only compresses the distances in the beginning of the chain (but leaves the remaining distances unchanged), and thereby causes stronger, but disordered couplings.
Although a stronger coupling increases the speed of the excitation transfer between two sites, for $v=0$ the efficiency of energy transfer, i.e.\ the sink population in the long-time limit, is less than that of a uniform chain without a pulse being present (solid line in figure~\ref{popGuided} (left)).
In fact, when the couplings at the beginning of the chain are stronger than between the remaining sites, they cause a localization of the excitation population at the beginning, due to the caused disorder in the couplings.
Therefore, only a relatively small fraction of the population succeeds to reach the sink, whereas the localized population is finally dissipated into the environment.
With this suppression, we once again face a distinct feature of coherent energy transport. In the classical, diffusive energy transfer, an increase of the hopping rates always causes an increase of the transfer rates, and consequently leads to a higher energy transfer efficiency.

If the pulse moves, it will sweep along the excitation that is localized within the pulse.
In the limit of very distant sites, a good strategy is to apply a strong pulse that couples only two sites and moves at the speed given by the Rabi frequency, such that after a full transfer to the second site, the pulse moves and couples site two and three.
For a pulse that is too fast, only part of the excitation will be transferred to the next site, whereas if it is too slow, part of the excitation oscillates back onto the first site again.
In the present case, we cannot neglect the coupling outside the pulse.
Instead, we need to coordinate the two processes of localizing the excitation within the moving pulse, and the wave-like coherent transfer in the uniformly coupled chain.
Figure~\ref{popGuided}~(left) shows that, for the given parameters and a chain length of 13~sites, the dominant contribution to the sink population due to the wave-like spreading of the excitation arrives between times 6--8 (in units of $1/J_0$), with small additions at later times.
The standing wave packet (dotted line) localizes the excitation at the beginning of the chain, but a fraction still succeeds to arrive in the sink at the same time as in the uniformly coupled chain.
The increase of the sink population starts earlier than without the pulse being present, because the pulse also slightly increases the effective coupling strength for the rest of the chain, which leads to a faster propagation of the excitation.
A wave packet moving with the speed of one site per time unit consequently arrives at the end of the chain after having passed 12 sites at time~$t=12$, where it leaves part of the excitation that it has swept along, resulting in an additional increase of the sink population.
At the optimal speed of $v_\text{opt}=2.53\pm0.01$, the wave packet arrives at about the same time as the excitation would in the uniformly coupled chain, hence effectively localizing the excitation in the front of the wave-like propagation.

The optimum speed is proportional to the coupling strength, i.e.\ it increases linearly with~$J_0$, because the Rabi frequency that is given by the coupling strength (up to units) is the speed at which an excitation moves between two sites during the Rabi oscillation. Therefore, the pulse has to match the speed given by~$J_0$. 

\begin{figure}
\centering
\includegraphics[width=0.5\textwidth]{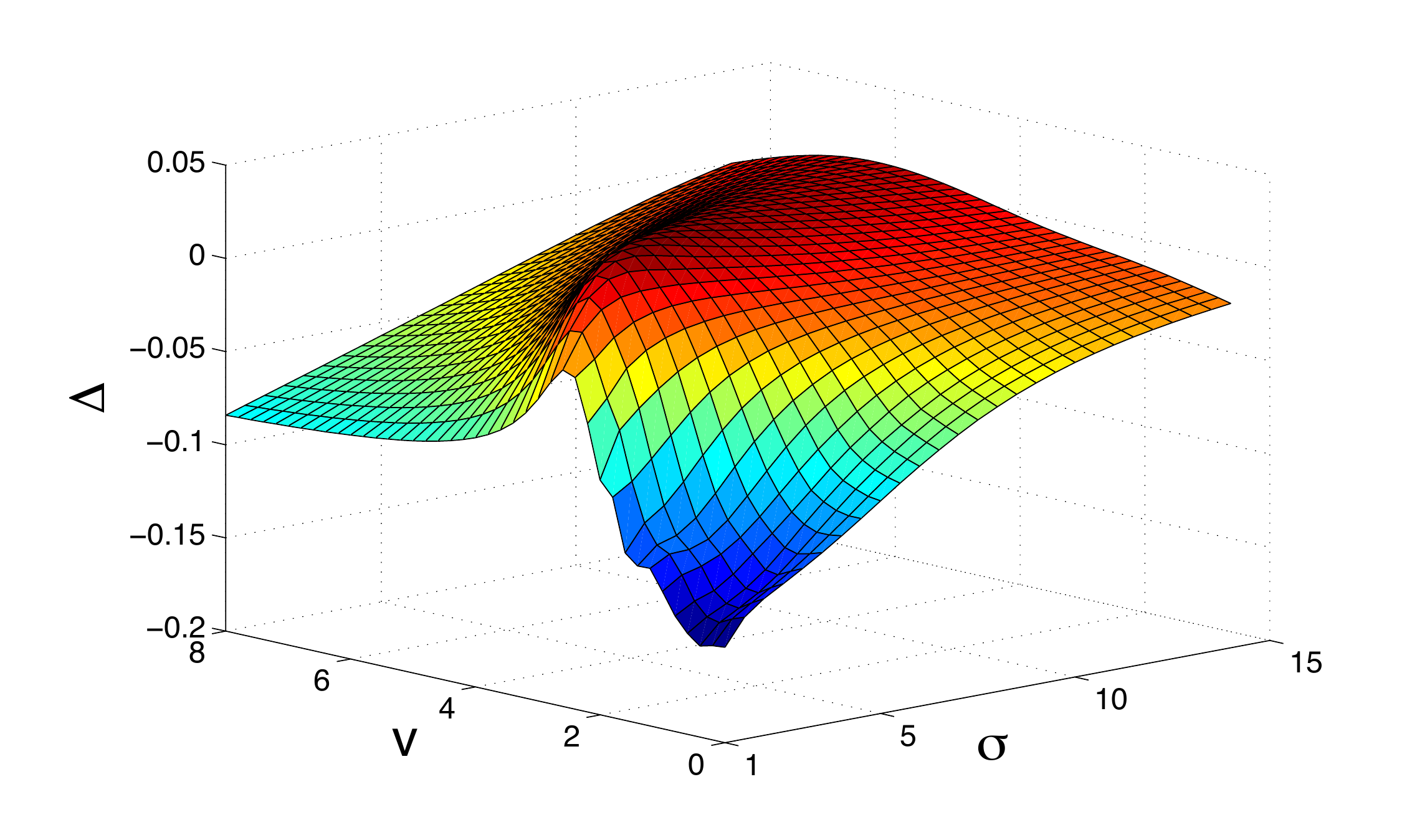}%
\includegraphics[width=0.5\textwidth]{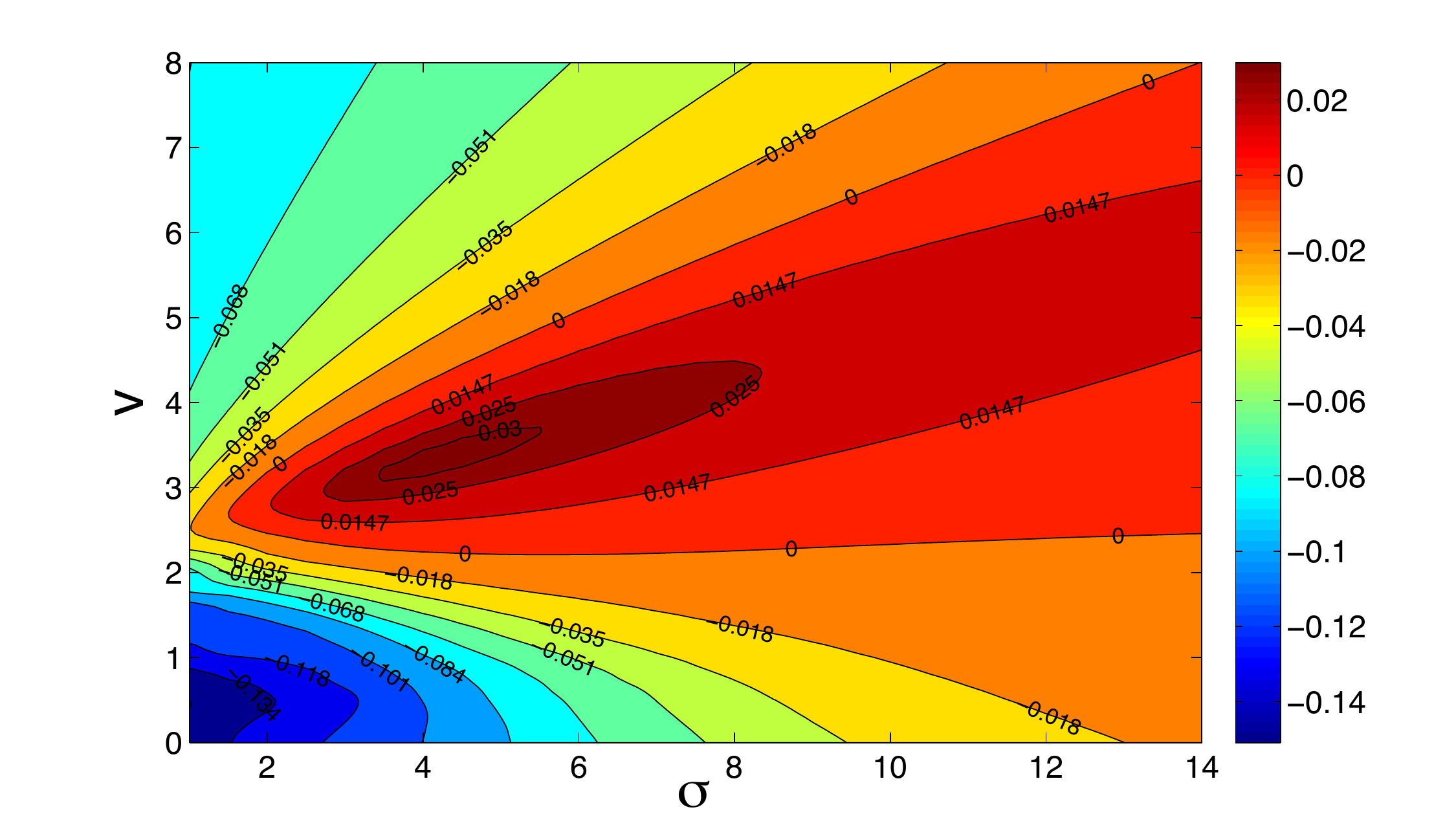}
\caption{Energy transfer enhancement for a Gaussian pulse over the uniformly coupled chain with $J_\text{max}$. Chain length is $N=13$, and $A=1/6$. Contours of the surface (left) are plotted on the right.}
\label{packetParameters}
\end{figure}

The simulation is summarized in figure~\ref{packetParameters}, which shows the increase of the sink population for a Gaussian pulse as compared to the uniformly coupled chain, $\Delta=P_\text{sink}(v,\sigma)-P_\text{sink}^\text{st}(J_\text{max})$, i.e.\ at a fixed maximum intensity of the pulse.
For small pulse widths and low velocities the gradient of coupling strengths changes quickly, and influences only few sites.
Therefore, the efficiency suffers most due to the strong localization of the excitation within the pulse.  
On the other hand, for large values of $\sigma$ the wave packet is broad, that is, it covers a larger region of the chain, and the gradient of coupling strengths changes more smoothly, which leads to less details in the velocity dependence.
In the limit of large pulse widths, the wave packet virtually covers the entire chain such that the reference case with $J_\text{max}$ is reached, and the efficiency gain approaches zero.
We also found that there is a global optimum ($v_\text{opt}=3.34\pm0.01$, $\sigma_\text{opt}=4.20\pm0.05$) where, at a given maximal intensity, the velocity and pulse width match best the coherent dynamics and achieve an increase of the transfer efficiency of more than $0.03$ over the static chain with maximal coupling strength. Note that the local optimum from figure~\ref{popGuided} with $\sigma=1$ corresponds to the maximum of the front-left cut of the surface in figure~\ref{packetParameters}~(left).

Finally, we also account for a possible dephasing, as caused by fast thermal noise, for example, which is formally implemented by adding the term~\eqref{dephasing} to the Lindblad equation.
The results are collected in the right panel of figure~\ref{popGuided}, and show that even under substantial dephasing the gain due to the presence of the pulse persists, and may even attain values in the classically forbidden region of enhancements $\Delta>0$ (not shown).

\section{Concluding remarks}

In this work, we demonstrate general features of energy transfer efficiency in mechanically oscillating systems, and relate our model to the biological scenarios of coherent energy transport in proteins.
Complementary to previous works, we focus on the driving of the excitation dynamics due to motion of the underlying molecular structure.
The time dependence of the inter-site coupling arises from its distance dependence together with relative motion of the sites, for example.
We generally find a motion-induced quantum enhancement of the excitation transfer efficiency over the equilibrium configuration of the linear chain, and furthermore over any static configuration that is met during the mechanical motion.
This effect distinguishes the quantum-coherent transport from the classical, diffusive transport.
If biological systems manage to utilize the modulation of coherent couplings by their motion, they can profit from the resulting enhancement of the transfer efficiency.
It would be interesting to extend our ideas beyond the linear setting to the enhancement of excitation transfer in light harvesting complexes in photosynthesis or light harvesting complexes in artificial devices.

\ack

Discussions with Leonor Cruzeiro and Victor Atanasov at the Workshop QuEBS 2009 in Lisbon, Portugal, are gratefully acknowledged.
We also thank J\"org Matysik for stimulating discussions and for bringing several references on transport in bacterial reaction centers to our attention.
This work was supported in part by the Austrian Science Fund (SFB FoQuS, J.M.C. through Lise Meitner Program) and the European Union (SCALA, NAMEQUAM)\@.
S.P.~acknowledges support from the UK EPSRC through the IRC-QIP.

\section*{References}

\end{document}